\documentclass[twocolumn,tighten,trackchanges,twocolappendix]{aastex701}

\usepackage{amsmath}

\newcommand{\oii}{\hbox{[O\sc ii]}}
\newcommand{\oiilong}{\hbox{[O\sc ii]\,3726,3729\AA}}
\newcommand{\oiialong}{\hbox{[O\sc ii]\,3726\AA}}
\newcommand{\oiiblong}{\hbox{[O\sc ii]\,3729\AA}}

\newcommand{\oiii}{\hbox{[O\sc iii]}}
\newcommand{\oiiilong}{\hbox{[O\sc iii]\,4959,5007\AA}}

\newcommand{\oiiiblong}{\hbox{[O\sc iii]\,5007\AA}}
\newcommand{\oi}{\hbox{[O\sc i]}}

\newcommand{\oialong}{\hbox{[O\sc i]\,6300\AA}}
\newcommand{\nii}{\hbox{[N\sc ii]}}
\newcommand{\niilong}{\hbox{[N\sc ii]\,6548,6583\AA}}

\newcommand{\niiblong}{\hbox{[N\sc ii]\,6583\AA}}
\newcommand{\sii}{\hbox{[S\sc ii]}}
\newcommand{\siilong}{\hbox{[S\sc ii]\,6716,6731\AA}}
\newcommand{\siialong}{\hbox{[S\sc ii]\,6716\AA}}
\newcommand{\siiblong}{\hbox{[S\sc ii]\,6731\AA}}

\newcommand{\ha}{H$\alpha$}
\newcommand{\hb}{H$\beta$}

\newcommand{\fluxcgs}{$\mathrm{erg}\,\mathrm{s}^{-1}\,\mathrm{cm}^{-2}$}
\newcommand{\lumcgs}{$\mathrm{erg}\,\mathrm{s}^{-1}$}

\newcommand{\msun}{$M_{\odot}$}

\newcommand{\lsun}{$L_{\odot}$}
\newcommand{\kms}{km\,s$^{-1}$}

\newcommand{\sfrunit}{\hbox{$M_{\odot}$\,yr$^{-1}$}}
\newcommand{\ccm}{cm$^{-3}$}

\newcommand{\todo}{\ifmmode \text{\color{purple}\Huge{\(\bullet\)}} \else {\color{purple}{\Huge$\bullet$}}\fi}
\newcommand{\finish}{\ifmmode \text{\color{blue}\Huge{\(\bullet\)}} \else {\color{blue}{\Huge$\bullet$}}\fi}

\shorttitle{A $z\sim0.4$ LRD-like galaxy: extended starburst with overmassive SMBH}
\shortauthors{Chen et al.}
\received{October 3, 2025}
\revised{December 18, 2025}
\accepted{January 7, 2026} %
\submitjournal{ApJ}

\begin{document}

\title{A $z\simeq0.4$ Galaxy Reflecting the high-redshift Little Red Dots: An Extended Starburst with an Overmassive Black Hole}

\correspondingauthor{Xiaoyang Chen}
\email{astro@xychen.me}

\author[0000-0003-2682-473X]{Xiaoyang Chen}
\affiliation{Frontier Research Institute for Interdisciplinary Sciences, Tohoku University, Sendai 980-8578, Japan}
\affiliation{Waseda Research Institute for Science and Engineering, Waseda University, 1-6-1 Nishi-Waseda, Shinjuku-ku Tokyo 169-8050, Japan}
\email{}

\author[0000-0002-4377-903X]{Kohei Ichikawa}
\affiliation{Frontier Research Institute for Interdisciplinary Sciences, Tohoku University, Sendai 980-8578, Japan}
\email{}

\author[0000-0002-2651-1701]{Masayuki Akiyama}
\affiliation{Astronomical Institute, Tohoku University, 6-3 Aramaki, Aoba-ku, Sendai, Miyagi 980-8578, Japan}
\email{}

\author[0000-0001-9840-4959]{Kohei Inayoshi}
\affiliation{Kavli Institute for Astronomy and Astrophysics, Peking University, 5 Yiheyuan Road, Haidian District, Beijing 100871, P. R. China}
\email{}

\author[0000-0002-7779-8677]{Akio K. Inoue}
\affiliation{Waseda Research Institute for Science and Engineering, Waseda University, 1-6-1 Nishi-Waseda, Shinjuku-ku Tokyo 169-8050, Japan}
\affiliation{Faculty of Science and Engineering, Waseda University, 1-6-1 Nishi-Waseda, Shinjuku-ku Tokyo 169-8050, Japan}
\email{}

\author[0000-0003-2984-6803]{Masafusa Onoue}
\affiliation{Waseda Institute for Advanced Study, Waseda University, 1-21-1, Nishi-Waseda, Shinjuku, Tokyo 169-0051, Japan}
\affiliation{Kavli Institute for the Physics and Mathematics of the Universe, UTIAS, Tokyo Institutes for Advanced Study, University of Tokyo, Chiba, 277-8583, Japan}
\email{}

\author[0000-0002-3531-7863]{Yoshiki Toba}
\affiliation{Department of Physical Sciences, Ritsumeikan University, 1-1-1 Noji-higashi, Kusatsu, Shiga 525-8577, Japan}
\affiliation{Research Center for Space and Cosmic Evolution, Ehime University, 2-5 Bunkyo-cho, Matsuyama, Ehime 790-8577, Japan}
\affiliation{Institute of Astronomy and Astrophysics, Academia Sinicam 11F of Astronomy-Mathematics Building, AS/NTU, No.1, Sec. 4, Roosevelt Rd, Taipei 106319, Taiwan, R.O.C.}
\email{}

\author[0000-0002-7051-1100]{Jorge Zavala}
\affiliation{Department of Astronomy, University of Massachusetts, LGRT-B 619E, 710 North Pleasant Street, Amherst, MA 01003-9305, United States}
\email{}

\author[0000-0002-5268-2221]{Tom Bakx}
\affiliation{Department of Space, Earth, \& Environment, Chalmers University of Technology, Chalmersplatsen 4, 412 96 Gothenburg, Sweden}
\email{}

\author[0000-0002-3866-9645]{Toshihiro Kawaguchi}
\affiliation{Graduate School of Science and Engineering, 3190 Gofuku, Toyama City, Toyama, 930-8555, Japan}
\email{}

\author[0000-0003-4814-0101]{Kianhong Lee}
\affiliation{Department of Physics, Graduate School of Science, Nagoya University, Furo, Chikusa, Nagoya, Aichi 464-8602, Japan}
\affiliation{Astronomical Institute, Tohoku University, 6-3 Aramaki, Aoba-ku, Sendai, Miyagi 980-8578, Japan}
\affiliation{National Astronomical Observatory of Japan, 2-21-1 Osawa, Mitaka, Tokyo 181-8588, Japan}
\email{}

\author[0000-0002-8299-0006]{Naoki Matsumoto}
\affiliation{Astronomical Institute, Tohoku University, 6-3 Aramaki, Aoba-ku, Sendai, Miyagi 980-8578, Japan}
\email{}

\author[0000-0003-2213-7983]{Bovornpratch Vijarnwannaluk}
\affiliation{Institute of Astronomy and Astrophysics, Academia Sinicam 11F of Astronomy-Mathematics Building, AS/NTU, No.1, Sec. 4, Roosevelt Rd, Taipei 106319, Taiwan, R.O.C.}
\affiliation{Astronomical Institute, Tohoku University, 6-3 Aramaki, Aoba-ku, Sendai, Miyagi 980-8578, Japan}
\email{}

\begin{abstract}
One of the most remarkable discoveries of JWST is a population of compact, red sources at $z>4$,
commonly referred to as Little Red Dots (LRDs).
Spectroscopic identifications reported that most LRDs are active galactic nuclei (AGNs),
which are preferentially found around $z\sim6$ 
and could imply a key phase in the formation and growth of black holes (BHs) in the early universe.
Photometric surveys at lower redshift have recently been carried out to trace their evolution across cosmic time,
and a small number of LRD-like galaxies have been spectroscopically identified at both Cosmic Noon and in the local universe.
Here we report the discovery of one of the lowest-$z$ LRD-like galaxies,
J204837.26-002437.2 (hereafter J2048) at $z=0.4330$,
using new Gemini-N/GMOS IFU observations combined with archival multi-band photometric SED data.
The GMOS data reveal extended blue emission from starburst with a star formation rate of 400 \sfrunit, 
together with an extended, highly fast ionized outflow.
This is the first spectroscopic confirmation of extended host emission and outflow in an LRD-like galaxy,
providing a unique laboratory for understanding the nature of their high-redshift counterparts.
Moreover, J2048 would host an extremely overmassive BH with a BH–to–stellar mass ratio of $\simeq60\%$, 
with the BH mass and host stellar mass estimated to be $10^{10.2}$ and $10^{10.4}$\,\msun, respectively. 
We discuss the origin and evolutionary fate of J2048,
and the implications that such low-$z$ analogs have for interpreting the properties of high-$z$ LRDs.
\end{abstract}

\keywords{Active galactic nuclei (16), Ultraluminous infrared galaxies (1735), Supermassive black holes (1663), AGN host galaxies (2017)}


\section{Introduction} \label{sec:intro}

The advent of JWST has revealed a previously unknown population of compact, red sources at $z > 4$, 
commonly referred to Little Red Dots \citep[LRDs; e.g.,][]{matthee2024}. 
These sources show a distinctive v-shaped spectral energy distribution (SED), 
characterized by a blue continuum in the rest-frame UV, a red continuum in the rest-frame optical bands, 
with a turnover near the Balmer break \citep{kocevski2023,kocevski2025,barro2024,furtak2023,greene2024,setton2024,wang2025}. 
Spectroscopic follow-up has confirmed that over 60\% of LRDs host broad Balmer emission lines, 
indicating the presence of active galactic nuclei (AGNs) with inferred black hole (BH) masses of $M_\mathrm{BH} 
\sim 10^{6-8}$~$M_\odot$ \citep{maiolino2024,greene2024,hviding2025}.

Despite their AGN signatures, most LRDs show weak near-IR (NIR) to mid-IR (MIR) emission and non-detections by ALMA bands, 
suggesting a lack of hot ($T\simeq1000$~K) and warm ($T\simeq300$~K) dust heated by the AGN torus 
\citep{leung2024,perezgonzalez2024,williams2024,akins2024,setton2025}. 
It is discussed that a more extended dust distribution ($>10$~pc) 
rather than typical AGN dusty region \citep[e.g.,][]{nenkova2008,honig2019,nikutta2021}, 
along with new constraints on the dust content in LRDs from multiwavelength data \citep{casey2025}, 
is required to explain the weak warm dust emission \citep{li2025,K.Chen2025}. 
In addition, most LRDs lack X-ray detections with the exception of a few sources \citep{kocevski2025}, 
indicating either significant obscuration or intrinsically weak corona emission, 
possibly related to a super-Eddington accretion phase \citep{ananna2024,yue2024,inayoshi2024a}.

The origin of the v-shaped SED, or in other words, the blue excess in the rest-frame UV SED, is also under debate. 
Either a star-forming host galaxy or scattered AGN light is considered as possible explanations of the blue excess \citep[e.g.,][]{leung2024}. 
Most LRDs are spatially unresolved with effective radii $\lesssim100$--300 pc \citep[e.g.,][]{akins2024,kokorev2024}. 
Such point-like morphologies can be due to the survey depth \citep[e.g.,][]{Billand2025,Rinaldi2025}
and the wavelength of the imaging band to probe this property (e.g., the spatial resolution of JWST/NIRCam is lower at longer wavelength).
Several studies report extended emissions in JWST's imaging observations in shorter wavelength bands (rest-frame UV), 
which support the explanation with blue, young stars in their host galaxies \citep[e.g.,][]{Rinaldi2024,Billand2025,CChen2025LRD,Zhuang2025}. 
However, due to the distance of the high-$z$ LRDs, the extended features are very faint, 
which prevent us from a detailed view of the host properties of LRDs.

It is suggested that a powerful outflow could have the potential to affect the evolution of LRDs.
\cite{Billand2025} proposed that outflows from AGNs or supernovae could be one of the mechanisms to enlarge the galaxy size
via expelling gas from the inner regions (the other scenarios are mergers and direct gas accretion from the environment).
The negative feedback effect of outflows in a LRD, e.g., suppression of SMBH growth and/or star formation 
by regulating the fueling gas accretion, has been recently discussed by \cite{wang2025}.
Outflow has been seen in several LRDs with JWST/NIRSpec observations 
with both emission lines \citep[e.g., \oiii,][]{Cooper2025,DEugenio2025} and absorptions \citep[e.g., He I,][]{Juodzbalis2024b,wang2025},
although the sample is very limited probably due to the faintness of outflow features. 

Another key aspect of LRDs is their unexpectedly high number density at $z > 4$, reaching $\sim10^{-4}$~Mpc$^{-3}$, 
at least two orders of magnitude higher \citep{kokorev2024,kocevski2025} 
than UV-selected quasars at comparable UV luminosities
\citep{akiyama2018,matsuoka2018,matsuoka2023,kulkarni2019,niida2020,he2024}. 
The reason on different densities of LRDs and quasars is under debate,
which is possibly due to selection bias of those surveys of LRDs and quasars, 
e.g., the current JWST survey area is not sufficient to find rare bright quasars.
The different densities can be also explained by the different evolutionary phases of black holes through cosmic time \citep{inayoshi2024,inayoshi2025}. 
Unlike quasars whose number density declines toward higher redshift, 
the LRD population appears to increase with redshift, suggesting that they may represent a distinct phase in black hole evolution, 
possibly corresponding to the earliest growth stage of supermassive BHs (SMBHs).
The inferred BH masses of LRDs, assuming that the Balmer lines are broadened by the Doppler effect of gas motion surrounding the BHs, are
roughly two orders of magnitude smaller than the typical ones in luminous quasars \citep{greene2024,he2024}. 
The inferred BH masses could be even lower if the broadening of Balmer lines is due to scattering of electrons and hydrogen atoms
\citep[e.g.,][]{Kokubo2024,Chang2025,Rusakov2025,Torralba2025}.

The current JWST surveys imply a sharp decline in the abundance of LRDs at $z<6$ \citep[e.g.,][]{inayoshi2024,inayoshi2025}.
In order to explore evolution of LRDs through cosmic time, 
several recent efforts have sought to identify LRD-like candidates\footnote{
	In this paper, we use the term ``LRDs'' to refer specifically to the high-redshift ($z \gtrsim 4$) population originally identified in JWST surveys. 
	We refer to lower-$z$ galaxies that exhibit similar phenomenological features, e.g., a characteristic v-shaped SED, as ``LRD-like candidates'' or ``LRD-like galaxies''. 
	This terminology is adopted to emphasize the differing environments and physical conditions across cosmic time, 
	as well as the possibility that these systems are not physically identical in nature; 
	for example, high-$z$ LRDs may correspond to an early growth phase of SMBHs in the young Universe.
	\citep[e.g.,][]{inayoshi2025}.
}
at lower-$z$ ($z<4$).
The lower-$z$ LRD-like candidates are usually identified 
with a v-shaped SED as shown in the high-$z$ LRDs
through photometric color selections, 
e.g., \citet[][]{Billand2025} and \citet[][]{kocevski2025} with JWST imaging surveys, 
Euclid surveys \citep[][]{bisigello2025}, 
and \citet[][]{ma2025} with the Subaru/HSC SSP survey.
\cite{bisigello2025} and \cite{ma2025} confirm the declined number density of the LRD-like galaxies as redshift decreases to $z\sim2$, 
though the values of the number densities are still diverse
probably due to the different selection criteria in the two studies
that can affect the SED shape constraints.
\cite{Billand2025} reports that the outskirt component in LRD-like galaxies could become more apparent with decreasing redshift due to the formation of host galaxies.
A small number of 
lower-z LRD-like galaxies
have been identified with spectroscopic observations
at Cosmic Noon \citep[$z\sim2$--3;][]{Noboriguchi2023,Juodzbalis2024b,ma2025,Stepney2024,Rinaldi2025,wang2025} and in the local universe \citep{LinR2025,Lin2025}. 
Utilizing observations with higher S/N and wider wavelength coverage, 
those lower-$z$ LRD-like galaxies
can be used as a window to witness the properties of the LRDs at $z>4$
by exhibiting more observational details
e.g., extended host properties \citep{Rinaldi2025} as well as detailed emission and absorption line properties \citep{Lin2025}. 

\begin{figure*}[ht!]
    \begin{center}
		\includegraphics[trim=0  0 -36 0, clip, width=0.9\textwidth]{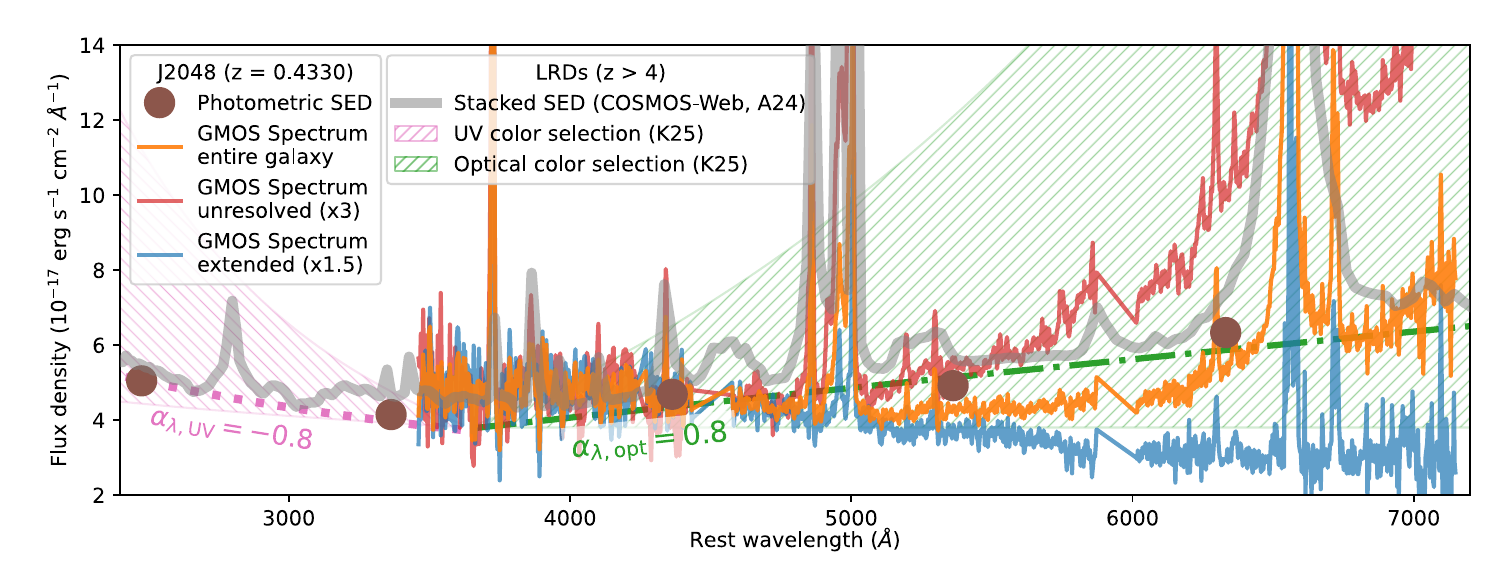}
    \end{center}
    \vspace{-20pt}
	\caption{
		GMOS spectra of J2048 integrated in the unresolved central region (red), the extended outskirt region (blue), and the entire galaxy (orange). 
		The unresolved and extended spectra are rescaled by factors of 3 and 1.5, respectively, to match the continuum flux of the entire galaxy at 3500--4500\,\AA.
		The photometric SED of J2048 in SDSS $ugriz$ bands are shown in brown filled circles.
		The best-fit power-law models for the photometric SED
		are shown in purple dotted (UV, $<3500$\,\AA) and green dash-dotted lines (optical, 4000--6500\,\AA), respectively.
		The purple and green hatched regions denote the color selection criteria of high-$z$ LRDs, 
		$-2.8<\alpha_\mathrm{\lambda,UV}<-0.4$ and $0<\alpha_\mathrm{\lambda,opt}<3$, respectively \citep{kocevski2025};
		which are normalized to the best-fit power-law models of J2048 (the dotted and dash-dotted lines) at the Balmer edge, 3646\AA. 
		The stacked spectrum of high-$z$ LRDs from the COSMOS-Web survey \citep{akins2024} is shown in the grey thick line,
		which is normalized to match the continuum flux of J2048 at 3500--4500\,\AA.
		J2048 has a LRD-like v-shaped SED at rest UV and optical wavelengths. 
		The red continuum ($>5000$\,\AA) as well as the broad \ha\ line profile ($\sim6500$\,\AA) are only exhibited in the unresolved central region. 
	}
	\label{fig:J2048_outskirt_comp}
\end{figure*}

\begin{figure*}[!ht]
    \begin{center}
		\includegraphics[trim=0 0 0 0, width=1.03\textwidth]{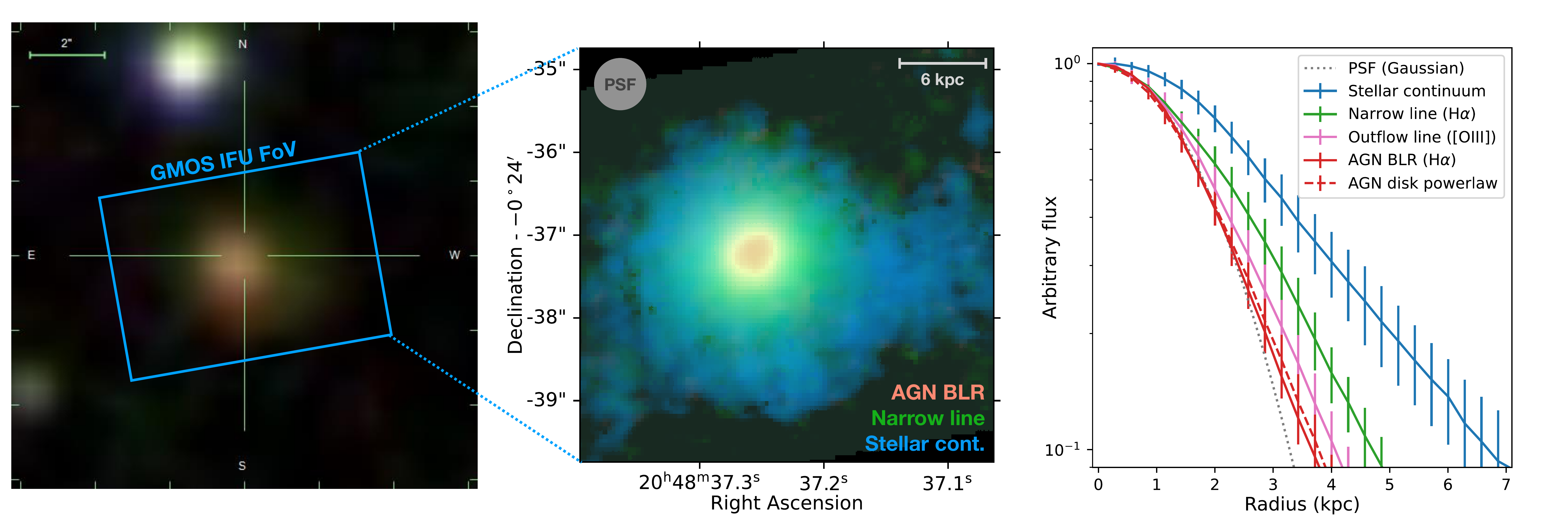}
    \end{center}
    \vspace{-10pt}
	\caption{
		Left: SDSS $gri$-bands composite image of J2048. 
		The field of view of the GMOS IFU observation is over-plotted as a blue rectangular. 
		Middle: Composite image of J2048 from the best spectral fitting results of GMOS IFU data. 
		The red, green, and blue colors denote the AGN BLR \ha\ line, narrow \ha\ line, and 
		young stellar continuum (integrated in rest frame 3500--4000\,\AA), respectively. 
		The grey circle shows the PSF FWHM of 0.65\arcsec.
		Right: Average radial profiles of young stellar continuum (blue; 3500--4000\,\AA), narrow \ha\ line (green), 
		outflow \oiii\ line (purple; the ``outflow line 1'' component in Figure \ref{fig:J2048_s1fit_line}), 
		AGN BLR (red solid) and disk power-law component (red dashed). 
		The profiles are arbitrarily normalized at the center. 
		An approximate PSF using a Gaussian profile with FWHM of 0.65\arcsec\ is shown in grey dotted line, 
		which can reproduce the bulk of the unresolved AGN BLR and power-law components. 
	}
	\label{fig:J2048_image}
\end{figure*}

In this paper, we present 
the discovery of one of the lowest-$z$ 
LRD-like galaxies, 
J204837.26-002437.2 (hereafter J2048) at\footnote{
	The systemic redshift is estimated with narrow \ha\ line. See Section \ref{subsec:method_ifu_spec} and \ref{subsec:result_narrow} for details. 
} $z=0.4330$,
by utilizing new Gemini-N/GMOS Integral field unit (IFU) observations as well as archived multi-band photometric SED.
J2048 is identified as a LRD-like galaxy since it has a v-shaped SED as that shown in high-$z$ LRDs, 
i.e., a blue continuum at rest UV and a red continuum at rest optical wavelengths. 
The approximate power-law indexes in rest UV ($<3500$\,\AA) and optical (4000--6500\,\AA)
are estimated to be $\alpha_\mathrm{\lambda,UV}=-0.8$ and $\alpha_\mathrm{\lambda,opt}=0.8$, respectively, 
using the photometric SED of J2048 in SDSS $ugriz$ bands (Figure \ref{fig:J2048_outskirt_comp}).
The power-law indexes are within the ranges of high-$z$ LRDs, e.g., $-2.8<\alpha_\mathrm{\lambda,UV}<-0.4$ and $0<\alpha_\mathrm{\lambda,opt}<3$
\citep[][see also the hatched regions in Figure \ref{fig:J2048_outskirt_comp}]{kocevski2025}. 
The GMOS IFU observations show that 
the red optical continuum ($>5000$\,\AA) as well as the broad \ha\ line profile ($\sim6500$\,\AA)
only appear in the unresolved central region, 
while the extended region exhibits a star-forming, blue optical continuum (Figure \ref{fig:J2048_outskirt_comp} and \ref{fig:J2048_image}). 
The result indicates a LRD-like nucleus surrounded by a star-forming host galaxy, 
which is the first time that a spatially extended, star-forming host galaxy is confirmed in a LRD-like system
with IFU spectroscopic observations.
The GMOS spectra also suggest a highly fast ionized outflow. 
These observations and results make J2048
a unique laboratory to understand the properties of those distant LRDs. 
We report the observation in Section \ref{sec:data} and analysis methods in Section \ref{sec:method}.
The results on AGN, host galaxy, and emission line properties are shown in Section \ref{sec:result}.
The comparison of observational properties between J2048 and high-$z$ LRDs 
is discussed in Section \ref{subsec:discuss_low_z_LRD};
the implications on high-$z$ LRDs from the observations of J2048 is discussed in Section \ref{subsec:discuss_implication};
the extremely overmassive SMBH is discussed in Section \ref{subsec:discuss_over_massive}. 
Cosmological parameters $H_0=$ 67.4 \kms\,Mpc$^{-1}$ and
$\Omega_\mathrm{M}=0.315$ \citep{Planck2020} are adopted throughout the paper. 


\section{Sample and Data} \label{sec:data}

\subsection{Sample Selection}
\label{subsec:selection}

J2048 is one of the ultra-luminous IR galaxies (ULIRGs) at intermediate redshifts ($0.1<z<1.0$),
which were selected 
from the cross-matched all-sky survey catalogs of AKARI (far-IR)
and Wide-field Infrared Survey Explorer (WISE, NIR and MIR)
as well as the Sloan Digital Sky Survey (SDSS, optical) spectroscopic catalog \citep{Chen2020}.
The parent sample is selected with the entire IR luminosity ($L_\text{1-1000\micron}$) higher than $10^{12}$\,\lsun.
Eight ULIRGs including J2048 exhibit highly fast ionized outflows ($>1500$ \kms) in their blueshifted, broad \oiiiblong{} line profiles. 
IFU follow-up observations with Gemini/GMOS and Subaru/FOCAS 
were conducted for the ULIRGs to understand the spatial extent and kinetic power of these extreme outflows.
Compared to the spatially-unresolved and fiber-aperture-limited spectroscopic data of the SDSS legacy survey,
the new IFU follow-ups provide spatially-resolved data for the entire galaxies
and cover longer wavelength ranges up to 1.05 \micron. 
A compact red continuum and a broad \ha\ line from the AGN Broad Line Region (BLR)
are newly detected in the Gemini/GMOS observation of J2048, 
which are not shown in the SDSS archived spectrum due to its limited spectral coverage. 
We show the observational details in the next subsection. 

\subsection{Gemini-N/GMOS IFU observations}
\label{subsec:Reduction_GMOS}

J2048 was observed by Gemini-N/GMOS in 2022 July
(ID: GN-2022A-Q-221, PI: Chen).
The observation was conducted with a total of 2.5 hours on-source exposure in two nights in the 2-slit IFU mode of GMOS. 
The 2-slit IFU mode has a field of view (FoV) of $5\arcsec\times7\arcsec$ with a pixel scale of $0.2\arcsec$. 
The FoV and position angle of the observation are shown in Figure \ref{fig:J2048_image}. 
The Point Spread Function (PSF) of the IFU data
has a Full Width Half Maximum (FWHM)
of $\sim0.65\arcsec$ in $i$-band ($\sim7500$\AA), 
which is estimated with stellar objects in the acquisition images of the observations. 
The grating R150 was utilized to achieve a wide wavelength coverage, 
i.e., from \oiialong\ line to \siiblong\ line (Figure \ref{fig:J2048_s1fit_line}), 
which provides a spectral resolution of R $\sim$ 1000 ($\sim$ 300 \kms) around 7000\AA.
The wavelength accuracy estimated with night sky emission lines is $\sim$ 50 \kms. 
The observation has a spectral pixel scale ($\Delta \lambda$) of 3.9\AA.

The GMOS data is reduced with the Gemini IRAF package \citep{GeminiIRAF}, e.g., to calibrate the wavelength, mask bad pixels, 
remove the scattered light, correct for the atmospheric dispersion, 
and subtract the night sky emissions. 
We follow the same reduction pipeline adopted in \cite{Chen2025}
and recommend readers to check the details in this paper.

\subsection{Archive observations}
\label{subsec:Reduction_archived}

We collect the multi-wavelength photometric observations of J2048 
in archives of the following telescopes and instruments:
SDSS ($u^\prime$ to $z^\prime$ bands), 
2MASS ($J$, $H$, and $K_s$ bands), WISE (3.4, 4.6, 12, and 22 \micron),
Spitzer IRAC (3.6, 4.5, 5.8, and 8 \micron) and MIPS (24 \micron).
The photometric fluxes are shown in Figure \ref{fig:J2048_s3fit_SED}. 
A simultaneous fitting for the new GMOS spectrum and the archived photometric data
is performed with the method described in Section \ref{subsec:method_s3fit}. 


\section{Analysis method} \label{sec:method}

\begin{figure*}[!ht]
    \begin{center}
		\includegraphics[trim=0 6 -36 0,  clip, width=0.7\textwidth]{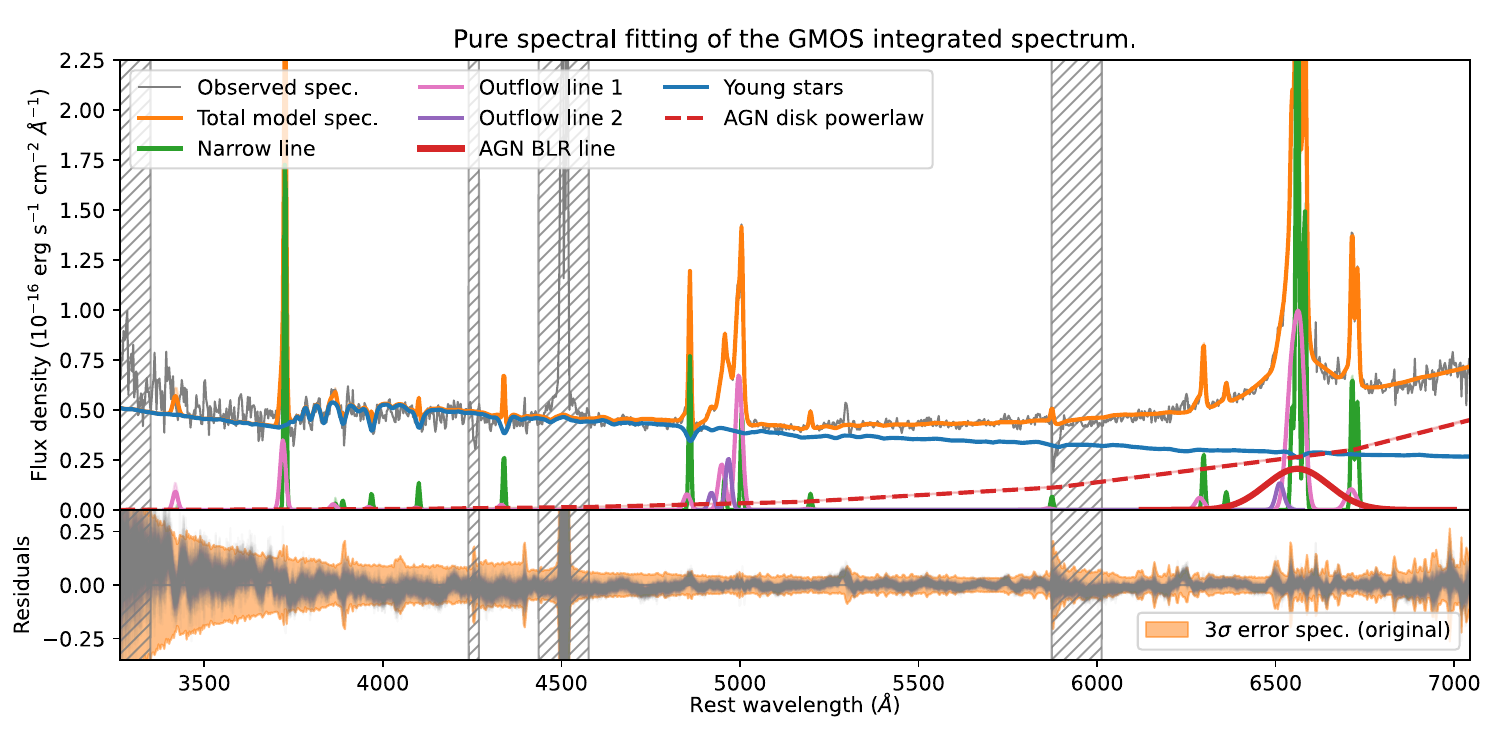}
		\includegraphics[trim=0 0 -36 18, clip, width=0.7\textwidth]{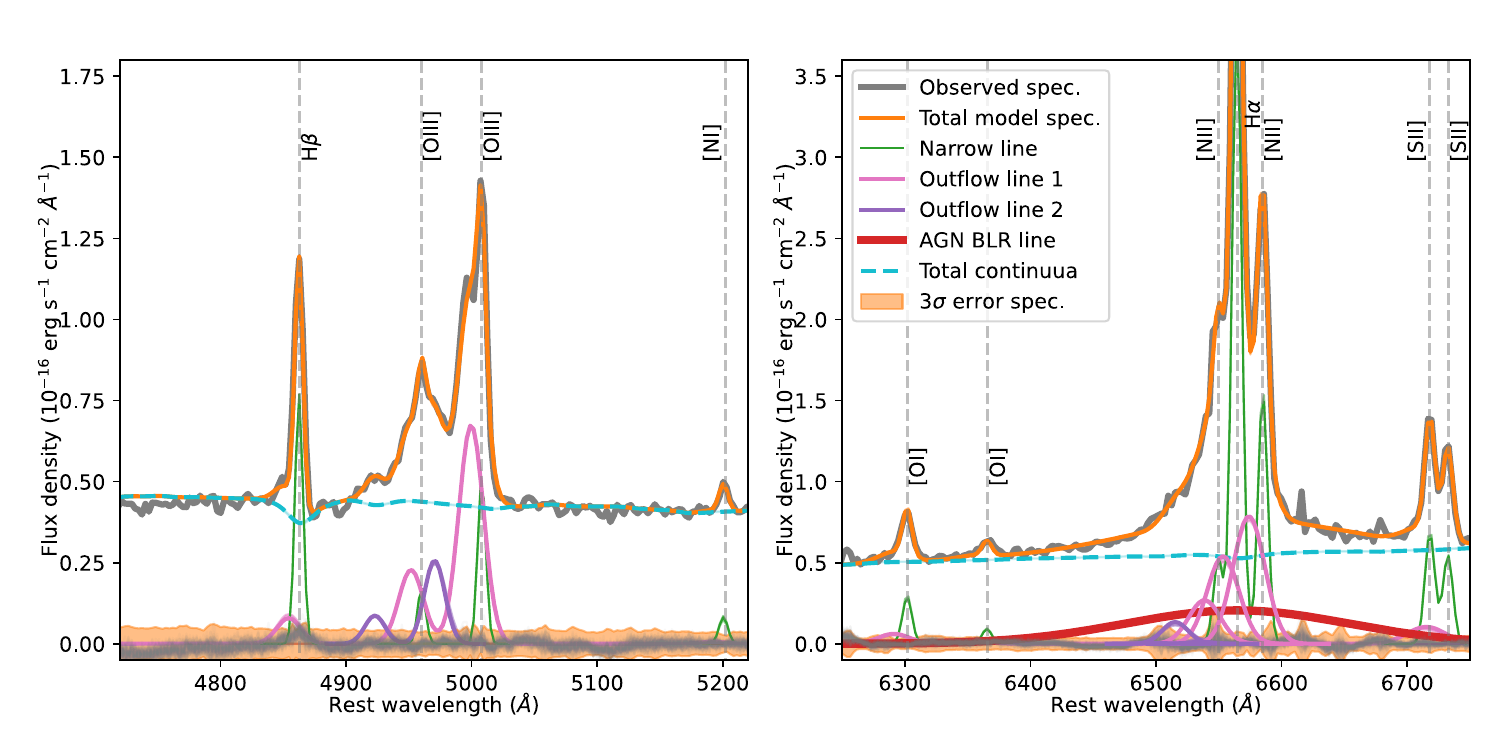}
    \end{center}
    \vspace{-20pt}
	\caption{
    	Best-fit results of the pure spectral fitting (Section \ref{subsec:method_ifu_spec}) for the GMOS spectrum of the entire galaxy
    	with fitting residuals (top and middle panels), 
    	as well as two zoomed-in regions around  \hb-\oiii\ (bottom-left) and \ha-\nii\ (bottom-right), respectively. 
    	The observed spectrum is shown in grey. 
    	The total model is shown in orange, with the young stellar continuum in blue and the AGN power-law component in red dashed lines. 
    	Emission lines include the narrow component (green), two broad blueshifted outflows (purple and violet), and the BLR H$\alpha$ line (thick red). 
    	Grey hatched areas are excluded due to poor data quality.
    	Note that in the top panel, the outflow profiles of the sum of \ha-\nii\ complex are shown for the sake of illustration. 
	}
	\label{fig:J2048_s1fit_line}
\end{figure*}

\begin{figure*}[!ht]
    \begin{center}
		\includegraphics[trim=0 16 -36 0, clip, width=0.7\textwidth]{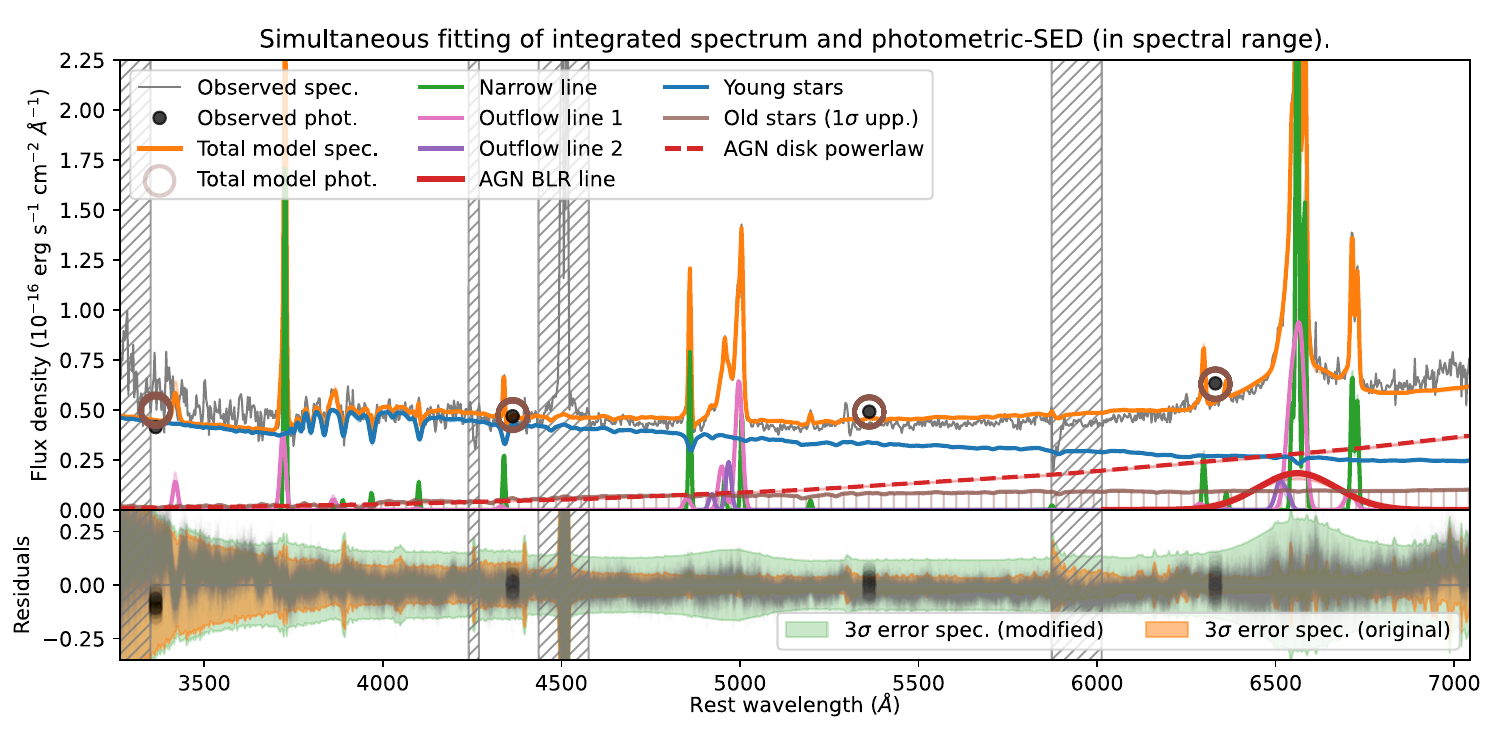}
		\includegraphics[trim=0 0 -36 12, clip, width=0.7\textwidth]{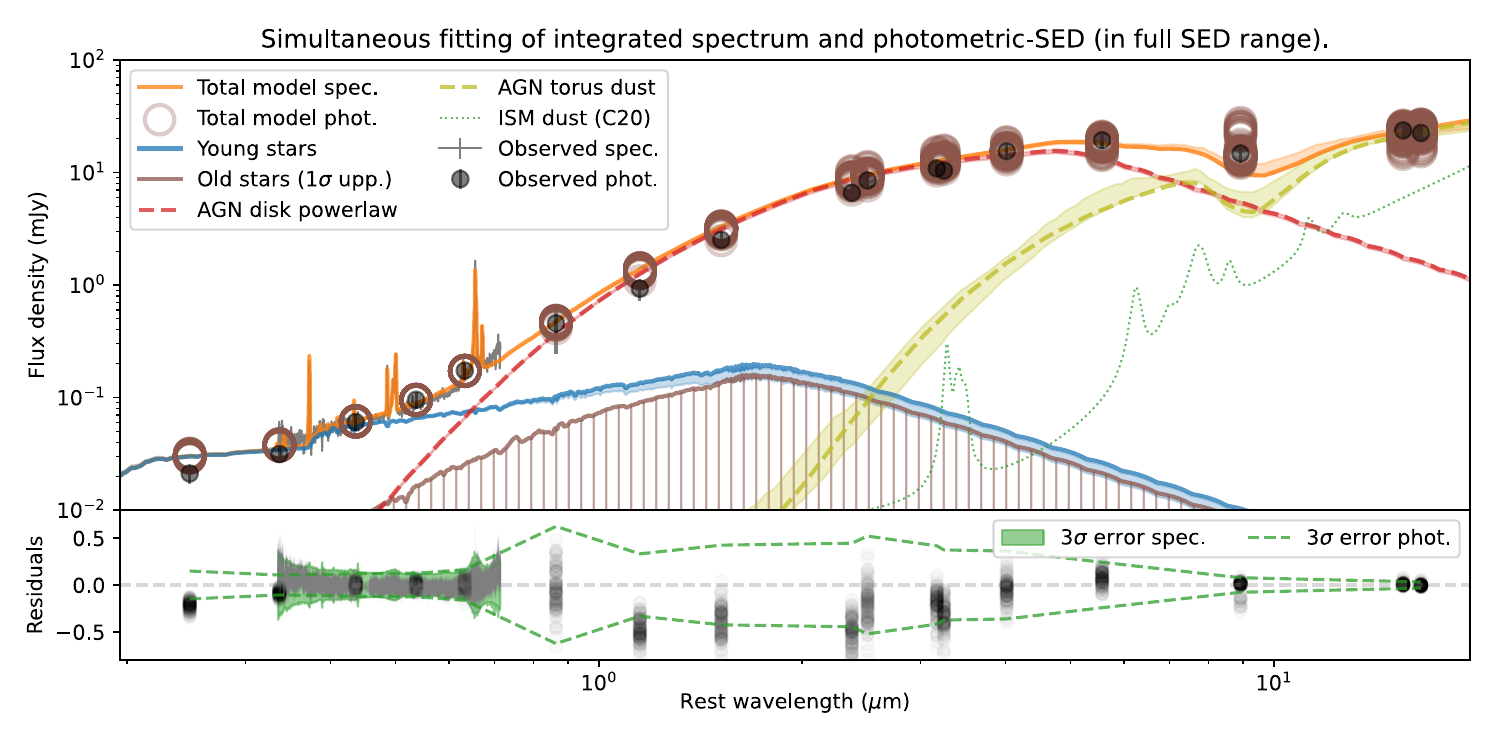}
    \end{center}
    \vspace{-24pt}
	\caption{
		Best-fit results of the simultaneous spectrum+SED fitting (Section \ref{subsec:method_s3fit})
		for observations of the entire galaxy
		in the wavelength range of the spectrum (top)
		and all used SED data (lower-middle; rest-frame 0.2--20 \micron)
		with their fitting residuals, respectively (upper-middle and bottom panels). 
		Black dots and brown open circles denote the photometric data and the best-fit models, respectively. 
		The legends of models are the same as those used in Figure \ref{fig:J2048_s1fit_line}.
		The AGN torus model is denoted in yellowgreen dashed curve. 
		Shaded areas show the 25\%–75\% uncertainty ranges from Monte Carlo simulation for each component.
		The unconstrained old stellar population is shown as a 1$\sigma$ upper limit (brown hatched; Section \ref{subsec:result_host}). 
		The best-fit ISM dust model of J2048 from \cite{Chen2020} is shown in green dotted line as a reference. 
		Note that all the residuals in the upper-middle and bottom panels are 
		shown in unit of $F_\lambda$ ($10^{-16}$ erg s$^{-1}$ cm$^{-2}$ \AA$^{-1}$) for the sake of illustration.
		The scatters reflect the fitting residuals of the original and mock data
		of the spectra (grey curves) and photometric points (black dots). 
		Modified errors (to reflect the calibration errors) are shown in green shadow region and dashed lines.  
	}
	\label{fig:J2048_s3fit_SED}
\end{figure*}

\subsection{Spectral fitting for the GMOS IFU spectra} \label{subsec:method_ifu_spec}

In order to determine the spatial extents of the host and AGN components, 
we perform per-pixel spectral fitting for the GMOS IFU data cube 
utilizing our fitting code\footnote{https://github.com/xychcz/S3Fit} \texttt{S$^3$Fit}
\citep{S3Fit,Chen2025}. 
The stellar continuum is fit utilizing 
an old stellar population (OSP) with an exponentially decayed star formation history (SFH),
and a young stellar population (YSP) with constant star formation rate (SFR). 
Both of OSP and YSP are built 
using the single stellar population (SSP) library of PopStar \citep{Millan-Irigoyen2021}
with the initial mass function (IMF) of \cite{Kroupa2001}. 
A power-law component ($f_\lambda \propto \lambda^{\alpha_\mathrm{PL}}$) is adopted to represent the radiation from the AGN accretion disk. 

The GMOS spectra covers a wide emission line range 
from \oii\ to \sii\ (Figure \ref{fig:J2048_s1fit_line}). 
Two Gaussian profiles are used to fit each line, 
with a narrow profile (FWHM $<$ 750 \kms) for the emission from dynamically quiescent gas, 
and a broad profile (750 $<$ FWHM $<$ 2500 \kms) for the emission from outflowing gas. 
A secondary outflow profile with 
velocity shift ($v_\mathrm{s}$) $<-1000$ \kms\ 
is required to describe the blueshifted wings of \oiii\ and \ha-\nii\ complex. 
Furthermore, 
we use an additional broad profile (FWHM $>$ 5000 \kms)
to model the gas emission in the AGN BLR. 
The BLR profile is only detected in \ha\ due to high dust extinction (see discussion in Section \ref{subsec:result_agn})
and thus disabled for other Hydrogen Balmer lines. 

We employ the following methods to reduce the degeneracy in the fitting. 
(1) A solar metallicity is adopted for stellar components, 
which is found as a typical value in local (U)LIRGs \citep[e.g.,][]{Perez2021}.
The extinction of stellar continuum ($A_{V,\mathrm{*}}$) is adopted as $0.5 A_{V,\mathrm{NL}}$, 
where $A_{V,\mathrm{NL}}$ is estimated from the Balmer decrement\footnote{
	The adopted intrinsic \ha/\hb\ ratio is 2.863, assuming Case B recombination
	with electron temperature of $10^4$\,K and electron density of $10^2$\,\ccm{} \citep{Hummer1987}.
	The adopted electron density is consistent to the value estimated from the narrow \siilong\ doublets
	of the integrated spectrum, $10^{2.1\pm0.1}$\,\ccm{}.
} of narrow lines
with the extinction law of \cite{Calzetti2000};
the factor 0.5 is an typical stellar-to-nebular extinction ratio in starburst galaxies \citep[e.g.,][]{Calzetti1997,Chen2025}. 
The effect of the assumed extinction ratio on estimation of properties (e.g., stellar mass) 
is discussed in Section \ref{subsec:result_host}.
(2) The AGN power-law index ($\alpha_\mathrm{PL}$) is fixed to $-1.7$, 
which is a typical value in composite quasar templates \citep[e.g.,][]{Francis1991,VandenBerk2001,Selsing2016,Temple2021}.
The effect of the assumed index on estimation of properties (e.g., bolometric luminosity) 
is discussed in Section \ref{subsec:result_agn}.
The extinction of the power-law continuum ($A_{V,\mathrm{PL}}$) is a free fitting parameter
following the extinction law of \cite{Calzetti2000}. 
(3) For either narrow or outflow line profiles, 
we tie the kinematic parameters, i.e., $v_\mathrm{s}$ and FWHM, of each emission line (e.g., \oii\ and \ha).
In addition, the $v_\mathrm{s}$ of BLR \ha\ line is tied to $v_\mathrm{s}$ of the narrow line component,
assuming that both of them reflect the systemic redshift.  
(4) The flux ratios of neighboring lines of the same ions
are fixed to their theoretical values, e.g., \oiiilong\ and \niilong \citep{Storey2000}; 
or the values calculated with \texttt{PyNeb} \citep{Luridiana2015}, e.g., \oiilong, 
with the electron density estimated from the \siilong\ doublets
and an assumed electron temperature of $10^4$\,K. 
We recommend readers to check the document\footnote{https://github.com/xychcz/S3Fit/blob/main/manuals\\/basic\_usage.md} of \texttt{S$^3$Fit}
for details of the line setup. 

Monte Carlo simulations are used to estimate the uncertainty of each fitting parameter. 
For each observed spectrum, we generate 100 mock spectra via adding random noise normalized by the measurement errors, 
and perform spectral fitting for the mock spectra. 
The standard deviations of best-fit parameters of the simulated spectra are used as the 
uncertainties of best-fit parameters of observed spectra. 

The map of AGN, stellar, and emission line components as well as the average radial profiles of them
are shown in Figure \ref{fig:J2048_image}. 
The best-fit results for the integrated spectrum of the entire galaxy
(i.e., in the region shown in the middle panel of Figure \ref{fig:J2048_image})
are shown in Figure \ref{fig:J2048_s1fit_line}. 
We discuss these results in details in Section \ref{sec:result}. 

\subsection{Simultaneous fitting for the GMOS integrated spectrum and multi-band photometric SED} \label{subsec:method_s3fit}

The fitting of the reddened AGN power-law component is sensitive to the spectrum at $\lambda > 6000$ \AA\ (rest frame), 
which can be affected by the cutoff and moderate S/N of GMOS spectral observation 
at the long wavelength end.
The fitting of the power-law model can in turn, affect other components such as the stellar continuum. 
In order to improve the fitting quality and 
to obtain better constraints on parameters (e.g., extinction of the AGN power-law), 
we employ a simultaneous fitting for 
the integrated GMOS spectrum 
and the archived multi-band photometric data from UV to MIR range. 

The simultaneous fitting is performed with the \texttt{S$^3$Fit} in the joint spectrum+SED fitting mode.
We keep using the same configurations of 
continuum and emission line models in Section \ref{subsec:method_ifu_spec}
and add the SKIRTor AGN torus model \citep{Stalevski2016}
to extend the fitting to NIR-MIR range\footnote{
	We do not use the interstellar medium (ISM) dust model in the fitting since it only contributes to $<10\%$
	of fluxes in NIR-MIR range for J2048 \citep{Chen2020}.
	The best-fit ISM model of \cite{Chen2020} is shown in Figure \ref{fig:J2048_s3fit_SED} as a reference. 
}. 
The AGN power-law continuum is declined at $\lambda>5\micron$ with $\alpha_\mathrm{PL}=-4$, 
which corresponds to the Rayleigh-Jeans branch of the thermal emission from the outer region of the accretion disk
\citep[e.g.,][]{Schartmann2005,Feltre2012,Stalevski2016}. 

\texttt{S$^3$Fit} works by minimizing the reduced chi-square value:
\begin{equation}
	\chi_{\nu}^2 = \frac{1}{\Sigma_i w_i - n_\mathrm{par}} \Sigma_i \left[w_i \left(\frac{D_i - M_i}{E_{i,\mathrm{mod}}}\right)^2 \right],
	\label{equ:chi_square}
\end{equation}
where $n_\mathrm{par}$ is the number of used model parameters.
$D_i$ and $M_i$ are the data and total model flux at each wavelength pixel and photometric band.
$w_i$ is the weight to balance the fitting of spectroscopic and photometric data.
We adopt $w=R \Delta \lambda/\lambda$ for the spectrum, where $R$ and $\Delta \lambda$ 
are the resolution and binning width of the spectrum; $w=1$ is adopted for photometric data in each band.
In order to account for calibration errors between different instruments,
modified errors, $E_{i,\mathrm{mod}}=\sqrt{E_{i,\mathrm{orig}}^2 + (10\%D_i)^2 }$, are employed,
where $E_{i,\mathrm{orig}}$ are the original measurement errors. 
The modified errors are also used to create mock data in the Monte Carlo estimation of fitting uncertainties as discussed in Section \ref{subsec:method_ifu_spec}. 
In the calculation of $E_{i,\mathrm{mod}}$, 
the original spectral data $D_i$ is convolved with a FWHM of $10^4$ \kms\ 
to avoid involving artifacts into the mock spectral data around bright emission lines. 

\subsection{Comparison and choice of results from the two fitting methods} \label{subsec:method_comp}

The best-fit results of 
the pure spectral fitting and 
the simultaneous spectrum+SED fitting 
for the spatially integrated data
are shown in Figure \ref{fig:J2048_s1fit_line} and \ref{fig:J2048_s3fit_SED}, respectively. 
For the sake of a direct comparison between results of the two methods,
we also extend the best-fit models of the pure spectral fitting to the full SED range in Figure \ref{fig:J2048_s1fit_test_exSED}. 
Due to the limited wavelength coverage of the GMOS spectrum, 
the AGN power-law model from the pure spectral fitting 
is not well constrained and cannot reproduce the observations in NIR range. 
Therefore, in the later analysis, 
we adopt the power-law properties (e.g., $A_{V,\mathrm{PL}}$ and luminosity)
from the simultaneous spectrum+SED fitting. 
We also use the best-fit stellar properties (e.g., mass and SFR)
from the simultaneous spectrum+SED fitting, 
since it can be affected by the continuum decomposition with the power-law model. 

On the other hand,
the values of emission line properties (e.g., velocities and fluxes) 
estimated from the two methods only have slight differences (Table \ref{tab:line_vel} and \ref{tab:line_flux}), 
except for the BLR \ha\ line, which is more sensitive to the fitting of the underlying continuum.
The results of the simultaneous spectrum+SED fitting show larger uncertainties than those of the pure spectral fitting
as the modified errors enlarge the fitting tolerance (e.g., Equation \ref{equ:chi_square})
and involve more scatters in the mock spectrum in Monte Carlo simulation, 
which are also reflected by the larger fitting residuals in Figure \ref{fig:J2048_s3fit_SED} (upper-middle panel). 
We adopt the results of emission lines from the pure spectral fitting throughout the paper
since they are derived with a better subtraction of the underlying continuum, 
and their uncertainties directly reflect the measurement errors of the GMOS spectrum. 
The effect of result of BLR \ha\ line on the estimation of black hole mass is discussed in Section \ref{subsec:result_agn}. 

\begin{table}[!ht]
	\caption{Velocity shifts\footnote{
		All of the $v_\mathrm{s}$ is relative to the systemic redshift $z=0.4330$ estimated from the $v_\mathrm{s}$ of narrow line. 
	} ($v_\mathrm{s}$) 
	and FWHM\footnote{
		All of the line widths are corrected for spectral resolution. 
	}
	of each emission line component\footnote{
	For each component, the upper value is derived from the pure spectral fitting for the GMOS integrated spectrum (Section \ref{subsec:method_ifu_spec}), 
	which is adopted in default throughout the paper;
	the lower value is derived from the simultaneous fitting for the spectrum and the photometric SED (Section \ref{subsec:method_s3fit}).
	} }
	\centering
	\begin{tabular}{r|DD}
		\hline
		\hline
		~ & \multicolumn2c{$v_\mathrm{s}$ (\kms)}  & \multicolumn2c{FWHM (\kms)}\\
		\hline
		\decimals
		Narrow line   		& 0.00$\pm$0.91   	& 357.41$\pm$4.30  \\
		~					& 0.08$\pm$3.99   	& 380.34$\pm$15.24  \\
		Outflow line 1   	& -504.11$\pm$17.92   	& 1360.93$\pm$27.13  \\
		~					& -496.30$\pm$54.07   	& 1350.00$\pm$104.24  \\
		Outflow line 2  	& -2246.06$\pm$27.90   	& 1113.65$\pm$61.13  \\
		~					& -2215.93$\pm$98.57   	& 1037.92$\pm$156.12  \\
		AGN BLR line   		& 0.00$\pm$0.91   	& 9517.29$\pm$350.20  \\
		~					& 0.08$\pm$3.99   	& 11853.13$\pm$1810.78  \\
		\hline
	\end{tabular}
	\label{tab:line_vel}
\end{table}

\begin{table*}[!ht]
	\caption{Fluxes of each component of main emission lines\footnote{
	The fluxes are in unit of $10^{-16}$ \fluxcgs, which are the observed values and not corrected for the extinction. 
	For each line component, the upper value is derived from the pure spectral fitting for the GMOS integrated spectrum (Section \ref{subsec:method_ifu_spec}), 
	which is adopted in default throughout the paper;
	the lower value is derived from the simultaneous fitting for the spectrum and the photometric SED (Section \ref{subsec:method_s3fit}).
	The component with ``-'' is not used in the fitting due to their faintness. 
	} }
	\tiny
	\centering
	\begin{tabular}{r|DDDDDDDDDD}
		\hline
		\hline
		~	  & \multicolumn2c{\oiialong} & \multicolumn2c{\oiiblong} & \multicolumn2c{\hb}  & \multicolumn2c{\oiiiblong}  
		      & \multicolumn2c{\oialong} & \multicolumn2c{\ha} & \multicolumn2c{\niiblong}  & \multicolumn2c{\siialong}  & \multicolumn2c{\siiblong} \\
		\hline
		\decimals
		Narrow line   	& 8.84$\pm$0.28    & 11.93$\pm$0.38    & 8.94$\pm$0.23    & 6.01$\pm$0.25    & 3.91$\pm$0.18    & 54.05$\pm$0.69    & 21.69$\pm$0.55    &  9.75$\pm$0.27    & 7.83$\pm$0.20  \\
		~			   	& 8.92$\pm$0.46    & 12.04$\pm$0.62    & 9.54$\pm$0.53    & 6.16$\pm$0.82    & 3.95$\pm$0.51    & 56.83$\pm$2.40    & 23.31$\pm$2.07    & 10.35$\pm$0.97    & 8.28$\pm$0.80  \\

		Outflow line 1  & 4.06$\pm$0.36    & 5.49$\pm$0.49    & 2.72$\pm$0.27    & 23.94$\pm$0.42    & 2.72$\pm$0.27    & 24.59$\pm$1.05    & 35.90$\pm$1.25    & 2.45$\pm$0.38    & 3.22$\pm$0.61  \\
		~			   	& 4.46$\pm$0.57    & 6.02$\pm$0.77    & 2.04$\pm$0.64    & 22.78$\pm$1.14    & 1.37$\pm$0.85    & 20.87$\pm$3.97    & 35.14$\pm$5.76    & 2.65$\pm$1.14    & 1.89$\pm$1.56 \\

		Outflow line 2 	& -    & -    & -    & 7.55$\pm$0.37    & -    & 4.99$\pm$0.75    & -\footnote{
																								We enable the ``Outflow line 2'' component for \nii\ doublets in the fitting, while they are not detected 
																								(Figure \ref{fig:J2048_s1fit_line}, bottom-right).}    & -    & -  \\
		~			   	& -    & -    & -    & 6.62$\pm$1.03    & -    & 4.58$\pm$2.12    & -    & -    & -  \\

		AGN BLR line   	& -  & - & -  & -    & -  & 65.59$\pm$1.87 & -  & - & - \\
		~			   	& -  & - & -  & -    & -  & 72.98$\pm$7.34 & -  & - & - \\
		\hline
	\end{tabular}
	\label{tab:line_flux}
\end{table*}

\subsection{Examination of AGN BLR and power-law components} \label{subsec:method_exam}

The AGN BLR \ha\ line and power-law continuum are related to the main finding of this work, 
i.e., a low-$z$ LRD-like galaxy with an overmassive SMBH (Section \ref{subsec:result_agn} and \ref{sec:discussion}). 
It is crucial to examine the detection of these components, 
i.e., if they are indeed required in the fitting, 
before detailed discussion of the results. 
We assess the significance of these components 
using the Akaike Information Criterion (AIC; \citealt{Akaike1974,Burnham2002}) 
and the Bayesian Information Criterion (BIC; \citealt{Schwarz1978,KassRaftery1995}, 
following recommendations for astrophysical models \citep[e.g.,][]{Takeuchi2000,Liddle2007}.
For a given dataset with noises following a Gaussian distribution, 
the criterion values can be calculated as
$\Delta\mathrm{AIC}=2\Delta n_\mathrm{par} + \Delta \chi^2$ and 
$\Delta\mathrm{BIC}=\Delta n_\mathrm{par} (\ln n_\mathrm{data})+ \Delta \chi^2$, 
where $\Delta n_\mathrm{par}$ is the difference of numbers of model parameters between two fitting configurations;
$n_\mathrm{data}$ is the number of data points;
$\Delta \chi^2$ is the difference of chi-square values between the two fitting results.
A netagive criterion value, e.g., $\Delta\mathrm{AIC}=\mathrm{AIC}_\mathrm{fit1}-\mathrm{AIC}_\mathrm{fit0}<0$, 
suggests the model configuration in fit1 is preferred. 

We perform the fitting without BLR \ha\ or power-law components
and calculate the $\Delta\mathrm{AIC}$ and $\Delta\mathrm{BIC}$, respectively.
The BLR \ha\ line is examined in rest-frame 6300--6800 \AA\ (e.g., Figure \ref{fig:J2048_s1fit_line}, bottom-right panel).
The pure spectral fitting results in 
$\Delta\mathrm{AIC}=\mathrm{AIC}_\mathrm{w.BLR}-\mathrm{AIC}_\mathrm{w/oBLR}=-1203$
and $\Delta\mathrm{BIC}=-1193$;
the simultaneous spectrum+SED fitting results in 
$\Delta\mathrm{AIC}=-48$ and $\Delta\mathrm{BIC}=-38$, 
all of the tests support the requirement of BLR \ha\ line in the fitting. 
The power-law continuum is examined in the full SED range (e.g., Figure \ref{fig:J2048_s3fit_SED}, lower-middle panel)
and we get the criterion of $\sim-1800$ for both of $\Delta\mathrm{AIC}$ and $\Delta\mathrm{BIC}$, 
which indicates that the power-law component is necessary to reflect the red bump in optical and NIR bands. 

In addition to the AIC and BIC tests of the AGN power-law component, 
we also check if the red bump in optical and NIR SED can be explained by other models, 
e.g., a highly obscured old stellar continuum (Case 1) or a hidden starburst continuum (Case 2). 
The best-fit results of Case 1 and 2 are shown in Figure \ref{fig:J2048_s3fit_test_redSSP}.
In Case 1, the obscured old stellar continuum with $A_V=5.7$ (brown dashed curve) can reproduce the red bump in SED (except for 1--3 \micron).
However, the derived stellar mass, $10^{13}$ \msun, 
exceeds that of the most massive galaxies 
by one order of magnitude \citep[e.g.,][]{Kormendy2013}.
It is also unlikely that the bulk of the old stellar population 
is fully obscured by dust in optical bands 
as they mainly locate in diffuse interstellar medium \citep[e.g.,][]{Charlot2000}.
Therefore, it is unlikely that the observed red continuum is mainly contributed by obscured old stars. 
In Case 2, in addition to the young stellar population observed in UV and optical bands (blue solid curve in Figure \ref{fig:J2048_s3fit_test_redSSP}),
the best-fit result shows an additional hidden starburst component with $A_V=7.2$ (cyan dotted curve).
The hidden starburst corresponds to an unphysically high SFR of $10^5$ \sfrunit\ in the recent $\sim10$ Myr,
suggesting that the possibility of a hidden starburst to explain the red bump can be ruled out. 

To summarize, 
both of the AGN BLR \ha\ line and power-law continuum are required to explain the observed spectrum and SED;
neither obscured old stars nor a hidden starburst could be a reasonable main contributor to the red bump in optical and NIR SED. 
Furthermore, we find that 
the power-law and BLR line components show consistent morphologies and luminosities, 
which are discussed in details in Section \ref{subsec:result_agn}. 


\section{Results} \label{sec:result}

\subsection{AGN properties} \label{subsec:result_agn}

\subsubsection{Optical continuum, broad \texorpdfstring{H$\alpha$}{H-alpha}, bolometric, and torus luminosities} \label{subsec:result_agn_lum}

The AGN is implied by the power-law continuum and the BLR \ha\ emission line.
Both of the two components share the almost same, unresolved spatial profile (Figure \ref{fig:J2048_image}, right panel), 
which can be described approximately using a Gaussian profile with FWHM of 0.65$\arcsec$.

Extinction correction is required to obtain the intrinsic luminosities of the power-law continuum and the BLR \ha\ line.
The simultaneous spectrum+SED fitting results in a best-fit extinction amount of $A_{V,\mathrm{PL}}=6.28\pm0.27$ for the power-law continuum.
Since the BLR \hb\ line is too faint to be detected (e.g., Figure \ref{fig:J2048_s1fit_line}, bottom-left panel), 
we assume the peak of the BLR \hb\ line with the 1$\sigma$ noise level
and the lower limit of $A_{V,\mathrm{BLR}}$ is then estimated\footnote{
	The adopted intrinsic \ha/\hb\ ratio is 2.615, assuming Case B recombination
	with electron temperature of $10^4$\,K and electron density of $10^9$\,\ccm{} \citep{Hummer1987}.
} to be 6.05, which is approximately consistent with $A_{V,\mathrm{PL}}=6.28$.
In the later analysis,  we adopt $A_{V,\mathrm{PL}}=6.28$ for correction of 
both of power-law continuum and the BLR \ha\ line
assuming that the BLR has the same amount of extinction
as that of the AGN accretion disk (i.e., the power-law component). 

The power-law continuum has a luminosity of $\lambda L_{5100}=10^{12.81\pm0.11}$ \lsun\ after extinction correction.
The intrinsic BLR \ha\ luminosity ($L_\mathrm{H\alpha,BLR}$) is estimated to be $10^{11.13\pm0.09}$ \lsun\ (or $10^{44.72\pm0.09}$ \lumcgs). 
It is reported that the AGN optical and BLR \ha\ luminosities has a tight log-linear relation over a wide range of luminosities and redshifts 
\citep[e.g.,][]{Greene2005,Shen2012,Jun2015}.
The ratio, $\log(L_\mathrm{H\alpha,BLR}/\lambda L_{5100})=-1.7$, is 0.4 dex lower than the empirical relation of \cite{Jun2015}, 
while the offset is within the distribution scatter of the sample to derive the relation \citep[e.g., Figure 11 of][]{Jun2015}.
The comparison indicates that the relation of the estimated $L_\mathrm{H\alpha,BLR}$ and $\lambda L_{5100}$ of J2048 is consistent with normal AGNs. 

In order to estimate the AGN bolometric luminosity ($L_\mathrm{AGN,bol}$), 
we employ an empirical optical bolometric correction, $L_\mathrm{AGN,bol}/\lambda L_{4400}=5.1$ \citep{Duras2020}.
The extinction-corrected $\lambda L_{4400}$ is estimated as $10^{12.85\pm0.11}$ \lsun\ from the simultaneous spectrum+SED fitting. 
$L_\mathrm{AGN,bol}$ is then estimated to be $10^{13.56\pm0.11}$ \lsun, 
where the uncertainty corresponds to the scatter of the best-fit $\lambda L_{4400}$.
The estimation of $\lambda L_{4400}$ and $L_\mathrm{AGN,bol}$ depends on extinction correction with $A_{V,\mathrm{PL}}$, 
which in turn is affected by the power-law index ($\alpha_\mathrm{PL}$).
In order to reduce the fitting degenaracy (Section \ref{subsec:method_ifu_spec}), 
in the default fitting $\alpha_\mathrm{PL}$ is fixed to $-1.7$, 
which is a typical value for composite quasar templates \citep[e.g.,][]{Temple2021}.
We also test fitting with $\alpha_\mathrm{PL}$ in a range from $-2.0$ to $-1.4$, 
which can lead to a variation of $\pm0.2$ dex in the estimated $\lambda L_{4400}$ and $L_\mathrm{AGN,bol}$. 
The empirical bolometric correction can involve an additional spread of $\sim0.3$ dex \citep{Duras2020}.
In the later discussion, we adopt the fiducial value, $L_\mathrm{AGN,bol}=10^{13.6\pm0.4}$ \lsun\ (Table \ref{tab:summary}),
with the uncertainties accounting for the scatters in the default best-fit results, the choice of assumed $\alpha_\mathrm{PL}$,
and the scatters in the empirical calibrations \citep{Duras2020}. 

The AGN torus IR luminosity ($L_\mathrm{AGN,torus}$) is estimated to be $10^{12.0\pm0.6}$ \lsun\
from the simultaneous spectrum+SED fitting. 
The ratio, $\log(L_\mathrm{AGN,torus}/L_\mathrm{AGN,bol})=-1.5$, is 0.4 dex lower than 
the typical value of AGNs at the high luminosity end \citep[$\sim10^{47}$ \lumcgs, e.g.,][]{Ichikawa2019_BASS}.
The ratio could imply that the covering factor of the dusty torus is relatively small, e.g., 0.2--0.3
(or a half-opening angle of 10$^\circ$--20$^\circ$), indicated by the energy balance of the absorbed emission of the central AGN
and the reemitted radiation of torus in the SKIRTor model \citep{Stalevski2016}. 

\begin{figure*}[!ht]
    \begin{center}
		\includegraphics[trim=16 6 -36 0,  clip, width=\textwidth]{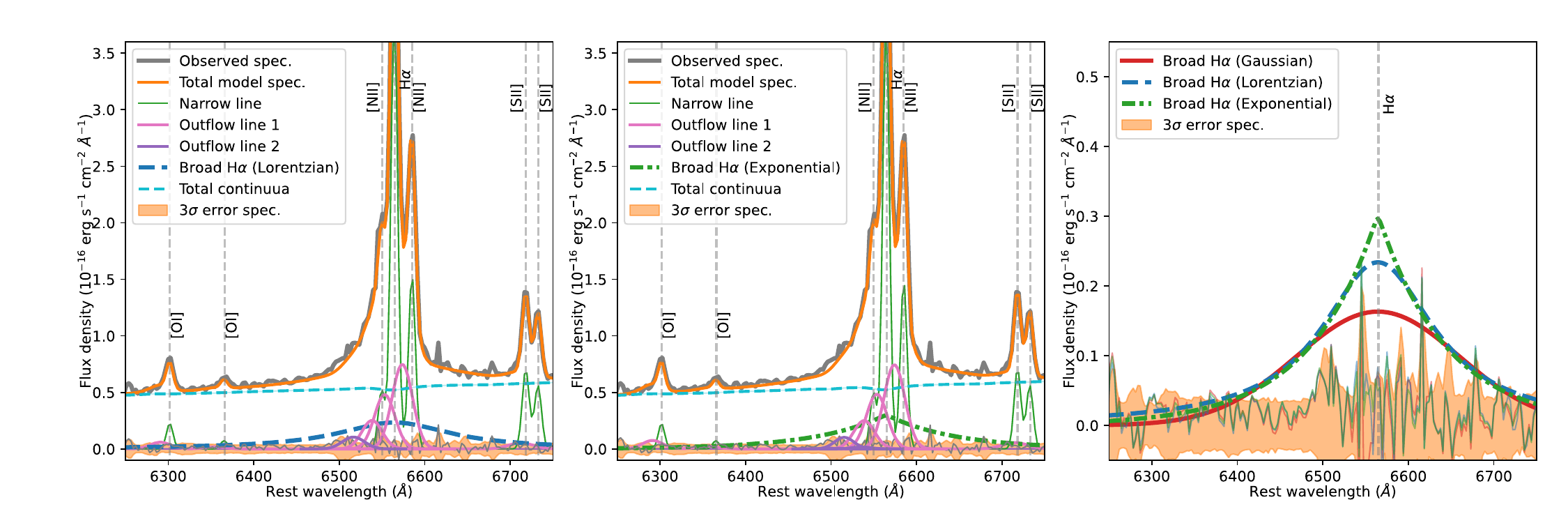}
    \end{center}
    \vspace{-20pt}
	\caption{
		Best-fit results of the integrated GMOS spectrum of the entire galaxy focusing the \ha-\nii\ wavelength range. 
		The broad \ha\ line is fit with a Lorentzian profile (blue dashed) and a double-side exponential profile (green dash-dotted)
		in the left and middle panels, respectively. 
		The other fitting configurations are the same as the default fitting (Section \ref{subsec:method_ifu_spec} and Figure \ref{fig:J2048_s1fit_line}). 
		The best-fit Gaussian (red solid curve; from the default fitting), Lorentzian, and exponential profiles of the broad \ha\ line are compared in the right panel;
		the fitting residuals are plotted in thin lines with the corresponding colors. 
	}
	\label{fig:J2048_s1fit_BLR_profile}
\end{figure*}

\subsubsection{SMBH mass} \label{subsec:result_agn_mbh}

The black hole mass ($M_\mathrm{BH}$) can be estimated with the AGN optical luminosity (e.g., $\lambda L_{5100}$) and 
the line width of the BLR \ha\ line following the empirical relation of \cite{Shen2012}:
\begin{equation} \label{equ:MBH}
\begin{split}
	\log \left( \frac{ M_\mathrm{BH} }{ M_\odot } \right) = 1.39 
	& + 0.56 \log \left( \frac{ \lambda L_{5100} }{ 10^{44}\,\mathrm{erg}\,\mathrm{s}^{-1} } \right) \\
	& + 1.87 \log \left( \frac{ \mathrm{FWHM_{H\alpha,BLR}} }{ \mathrm{km}\,\mathrm{s}^{-1} } \right).
\end{split}
\end{equation}
The FWHM of the BLR \ha\ line is estimated to be $9520\pm350$ \kms\ from the 
pure spectral fitting of the integrated spectrum
(Figure \ref{fig:J2048_s1fit_line} and Table \ref{tab:line_vel}). 
The $M_\mathrm{BH}$ is then derived as $10^{10.17\pm0.07}$ \msun.
The shown uncertainties above corresponds to the scatters of $\lambda L_{5100}$ and $\mathrm{FWHM_{H\alpha,BLR}}$
in the default fitting. 

The estimation of $M_\mathrm{BH}$ depends on the estimation of $\lambda L_{5100}$ and $\mathrm{FWHM_{H\alpha,BLR}}$, 
which can be affected by several factors. 
As discussed above in Section \ref{subsec:result_agn_lum}, 
different assumptions of $\alpha_\mathrm{PL}$ from $-2.0$ to $-1.4$
can involve a spread of $\pm0.2$ dex in the estimated $\lambda L_{5100}$, 
which corresponds to $\pm0.1$ dex in the derived $M_\mathrm{BH}$.
The line width of BLR \ha\ can be influenced by 
the choice of the fitting strategy (pure spectral fitting or joint spectrum+SED fitting; Section \ref{subsec:method_comp}), 
and the choice of line profiles to fit the BLR \ha\ line. 
If we adopt $\mathrm{FWHM_{H\alpha,BLR}}=11850$ \kms\ from the spectrum+SED fitting (Table \ref{tab:line_vel}),
the $M_\mathrm{BH}$ is estimated to be $10^{10.4}$ \msun.
In the default fitting, we fit the BLR \ha\ line with a Gaussian profile (Figure \ref{fig:J2048_s1fit_line}).
If we fit the BLR \ha\ line utilizing a Lorentzian profile \citep[e.g.,][]{Sulentic2002,Kollatschny2013,Rakshit2017},
the best-fit $\mathrm{FWHM_{H\alpha,BLR}}$ is 7350 \kms\ and the $M_\mathrm{BH}$ is estimated to be $10^{10.0}$ \msun.
There is no significant difference of fitting quality, e.g., the residuals compared in the right panel of Figure \ref{fig:J2048_s1fit_BLR_profile},
between fitting with Gaussian and Lorentzian profiles. 

In the later discussion, we adopt the fiducial value, $M_\mathrm{BH}=10^{10.2\pm0.3}$ \msun\ (Table \ref{tab:summary}), 
which is based on the default pure spectral fitting with a Gaussian BLR profile;
the shown uncertainty account for the scatters in the default best-fit results ($\pm0.1$ dex), 
the spreads due to the assumptions of $\alpha_\mathrm{PL}$ ($\pm0.1$ dex),
the choices of fitting strategy and line profiles ($\pm0.2$ dex), 
as well as the scatters in the empirical calibrations ($\sim0.1$ dex; \citealt{Shen2012}). 
Combining the estimated $L_\mathrm{AGN,bol}$ and $M_\mathrm{BH}$, 
the Eddington ratio of the AGN ($\lambda_\mathrm{Edd}$) is calculated to be $0.08\pm0.01$, 
suggesting the SMBH of J2048 is still growing via rapid accretion. 
The value is close to typical $\lambda_\mathrm{Edd}$ of high-$z$ LRDs, 0.1--0.4 \citep[e.g.,][]{greene2024,Kokorev2023}.

Finally we note that the above estimation is based on the assumption that 
the BLR \ha\ line is predominantly broadened by the Doppler effect of bulk gas motion 
gravitationally bound in the vicinity of the SMBH \citep[e.g.,][]{Harikane2023,maiolino2024,kocevski2025}.
Several recent studies of LRDs argued that the broadening can be due to 
Raman scattering or turbulence that leads to a Lorentzian profile,  
or Thomson scattering that leads to a double-sided exponential profile \citep[e.g.,][]{Kokubo2024,Chang2025,Torralba2025,Rusakov2025}.
The fitting with a Gaussian, a Lorentzian, and an exponential profile to the BLR \ha\ line are compared in Figure \ref{fig:J2048_s1fit_BLR_profile}.
Since there are no significant differences in the fitting quality (e.g., the fitting residuals),
the possibility of a scattering or turbulence dominated broadening cannot be ruled out. 
A detailed discussion on those broadening mechanisms is out of the scope of the paper. 
In the cases of non-Doppler broadening, we estimate $M_\mathrm{BH}$ assuming an Eddington accretion ($\lambda_\mathrm{Edd}=1$; e.g., \citealt{Torralba2025}), 
which gives $M_\mathrm{BH}=10^{9.1}$ \msun.
We consider the value as the most conservative estimation of $M_\mathrm{BH}$ of J2048. 

\begin{table}[!ht]
	\caption{
		Main properties of J2048 ($z=0.4330$).
	}
	\vspace{-4mm}
	\centering
	\begin{tabular}{l|l}
		\hline
		\hline
		AGN (Section \ref{subsec:result_agn}) & ~ \\
		Optical luminosity ($\lambda L_{5100}$, in log \lsun)   	& $12.8\pm0.1$ 		\\
		Torus luminosity ($L_\mathrm{AGN,torus}$, in log \lsun)   	& $12.0\pm0.6$ 		\\
		Bolometric luminosity ($L_\mathrm{AGN,bol}$, in log \lsun)  & $13.6\pm0.4$ 		\\
		SMBH mass ($M_\mathrm{BH}$, in log \msun)  					& $10.2\pm0.3$      \\
		Eddington ratio ($\lambda_\mathrm{Edd}$)					& $0.08\pm0.01$   	\\
		BH-to-host mass ratio ($M_\mathrm{BH}/M_\mathrm{*}$)		& 60\%   	        \\
		\hline
		Host galaxy (Section \ref{subsec:result_host} and \ref{subsec:result_narrow}) & ~ \\
		Young population mass ($M_\mathrm{YSP}$, in log \msun)  	& $9.9\pm0.1$  		  \\
		Total stellar mass (fiducial $M_\mathrm{*}$, in log \msun)  & $10.4\pm0.6$ \\
		SFR based on continuum (in \sfrunit)   		        		& $400\pm60$   	   	\\
		SFR based on narrow \ha\ (in \sfrunit)   		       		& $140\pm10$   	    \\
		\hline
		Outflow (Section \ref{subsec:result_outflow}) & ~ \\
		Maximum velocity ($v_\mathrm{out}$, in \kms)   				& $2070\pm40$   	\\
		Mass-loss rate ($\dot{M}_\mathrm{out}$, in \sfrunit)  		& 160   	 		\\
		Kinetic power ($\dot{E}_\mathrm{k,out}$, in log \lumcgs)  	& 44.3   			\\
		Kinetic coupling facto ($\dot{E}_\mathrm{k,out}/L_\mathrm{AGN,bol}$) & 0.2\%   	 \\
		\hline
	\end{tabular}
	\label{tab:summary}
\end{table}

\subsection{Host galaxy properties} \label{subsec:result_host}

Since the AGN power-law continuum emitting from the accretion disk is highly obscured in UV and optical bands, 
the spectrum of J2048 at rest-frame $\lambda<6000$\,\AA\ 
is dominated by stellar continuum emitted by young stars even in the unresolved central region ($r<1.8$\,kpc; Figure \ref{fig:J2048_outskirt_s1fit}, top panel). 
In the outskirt region ($r>2$\,kpc) where the AGN components are not detected, 
the full observed spectrum in dominated by stellar light (Figure \ref{fig:J2048_outskirt_s1fit}, lower-middle panel). 
After correction for PSF using a Gaussian profile with FWHM of $0.65\arcsec$, 
the average effective radius (i.e., the radius enclosing half of the entire flux) is estimated to be 3 kpc.
The stellar disk is elongated to an effective radius of 7 kpc towards the west, 
showing a spiral arm-like structure in the direction (Figure \ref{fig:J2048_image}, middle panel). 

The best-fit results of the simultaneous spectrum+SED fitting suggest a total mass ($M_\mathrm{YSP}$) 
of $10^{9.9\pm0.1}$ \msun\ for the young stellar population, 
which was formed in the recent $\sim20$ Myr 
and corresponds to a mean SFR of $400\pm60$ \sfrunit\ assuming a SFH with constant SFR. 
As noted in Section \ref{sec:method}, 
in order to reduce the degeneracy in the fitting, 
the extinction of stellar continuum ($A_{V,\mathrm{*}}$) is adopted as $0.5 A_{V,\mathrm{NL}}$, 
where $A_{V,\mathrm{NL}}$ is estimated from the Balmer decrement of narrow lines
and the factor 0.5 is an empirical stellar-to-nebular extinction ratio in ULIRGs \citep[e.g.,][]{Chen2025}. 
The estimation of $M_\mathrm{YSP}$ and SFR can be affected by the adopted extinction ratio.
We test the fitting with the extinction ratio in a range of 0.1--0.7
with the corresponding $A_{V,\mathrm{*}}$ varying from 0.2 to 1.5;
the fitting quality is poor with an extinction ratio over 0.7.
The derived $M_\mathrm{YSP}$ varies from $10^{9.3}$ to $10^{10.1}$ \msun\
with the SFR from 100 to 600 \sfrunit\ in the test fits, 
which suggest the uncertainty ranges related to the assumed extinction ratio. 
The SFR can be also estimated independently with the IR luminosity emitted from ISM dust heated by stellar light.
Utilizing $L_\mathrm{IR,SF}=10^{12.5}$ \lsun\ \citep{Chen2020}
and the empirical function \citep{Calzetti2013} under the same conditions\footnote{
	SFR (in \sfrunit) can be estimated as $3.7\times10^{44}$ $L_\mathrm{IR,SF}$ (in \lumcgs)
	for a young stellar population formed in 10 Myr
	with an IMF of \cite{Kroupa2001} and a stellar mass range of 0.1--100 \msun\ \citep{Calzetti2013}. 
}, the IR-based SFR is estimated to be 450 \sfrunit, 
which is consistent with the optical-based SFR from the default fitting result with a factor of 1.1. 

The mass of the old stellar population ($M_\mathrm{OSP}$) is required to obtain the total stellar mass of the galaxy\footnote{
	Due to the moderate spectral resolution and S/N of stellar absorption features, 
	in this paper we do not use the stellar dispersion to estimate the dynamical mass. 
}.
Stellar radiation from old stars typically peaks at NIR range (e.g., 1--3 \micron), 
however, the archived NIR observations of J2048 is dominated by the reddened AGN power-law component, 
making it hard to directly estimate the mass of the old population. 
In order to estimate the upper limit of $M_\mathrm{OSP}$,
we assume an oldest population (i.e., with the highest mass-to-light ratio)
that was formed at $z=10$ with an exponential timescale of 100 Myr. 
The spectrum+SED fitting is repeated by scanning a series of $M_\mathrm{OSP}$ values.
The 1$\sigma$ upper limit on $M_\mathrm{OSP}$ (i.e., the one-sided 84.1\% confidence interval)
is derived to be $10^{11.0}$ \msun, 
which is defined as the value for which the profile-likelihood increases by $\Delta \chi_{\nu}^2=1.35$ 
\citep[e.g.,][]{Bevington2003,Wall2012}
from the best-fit case that shows zero OSP component.
The OSP SED corresponding to the 1$\sigma$ upper limit is shown as brown hatched regions in Figure \ref{fig:J2048_s3fit_SED}.

With the above analyses, the total stellar mass ($M_\mathrm{*}$) is constrained in a range 
from $10^{9.9}$ \msun\ (best-fit $M_\mathrm{YSP}$)
to $10^{11.0}$ \msun\ ($M_\mathrm{OSP}$ upper limit). 
In later discussion, we adopt a fiducial value of $M_\mathrm{*}=10^{10.4\pm0.6}$ \msun, 
which is estimated utilizing the average mass fraction of YSP, $0.3\pm0.2$, 
of 149 ULIRGs at $0.1<z<1$ \citep{Chen2020};
the shown uncertainty of $M_\mathrm{*}$ corresponds to the spread between the best-fit $M_\mathrm{YSP}$ and the upper limit of $M_\mathrm{OSP}$.
The fiducial $M_\mathrm{*}$ value
corresponds to a specific SFR of $10^{-7.7}$\,yr$^{-1}$, 
which is over one order of magnitude higher than the specific SFR of star forming main sequence ($\lesssim10^{-9}$\,yr$^{-1}$, \citealt{Peng2014})
and suggests a vigorous starburst phase. 

\subsection{Extended narrow emission lines} \label{subsec:result_narrow}

J2048 has an spatially extended narrow emission line region. 
The map and mean radial profile of narrow \ha\ line is shown in Figure \ref{fig:J2048_image}, 
which correspond to an effective radius of 2.5 kpc after correction for PSF blurring. 
The FWHM of the narrow line component is estimated to be $357\pm4$ \kms\ using the integrated spectrum
after correction for the spectral resolution (Table \ref{tab:line_vel}). 
Table \ref{tab:line_bpt} lists the flux ratios of emission lines that are used to determine the ionization energy source.
It is indicated that the extended narrow lines are ionized by stellar light of young stars
with the BPT diagnostics \citep{Baldwin1981,Kewley2001,Kauffmann2003}.
The corresponding instantaneous SFR estimated using the 
extinction-corrected fluxes of the
narrow \ha\ or \oii\ lines
are $140\pm10$ or $100\pm20$ \sfrunit, respectively, 
utilizing the empirical functions of \cite{Calzetti2013} and \cite{Kennicutt2009}. 

\begin{table}[!ht]
	\caption{Flux ratios of narrow and outflow emission lines.}
	\centering
	\begin{tabular}{r|DD}
		\hline
		\hline
		Ratio & \multicolumn2c{Narrow line} 
		& \multicolumn2c{Outflow line 1}\\
		\hline
		\decimals
		log(\oiii/\hb)   	& -0.17$\pm$0.02   	& 0.94$\pm$0.04  \\
		log(\nii/\ha)   	& -0.40$\pm$0.01   	& 0.16$\pm$0.02  \\
		log(\oi/\ha)   	    & -1.14$\pm$0.02   	& -0.96$\pm$0.05  \\
		log(\sii/\ha)   	& -0.49$\pm$0.01   	& -0.64$\pm$0.05  \\
		\hline
	\end{tabular}
	\label{tab:line_bpt}
\end{table}

\subsection{Fast and powerful ionized outflow} \label{subsec:result_outflow}

J2048 possesses a fast ionized outflow, 
which is indicated by the blueshifted, broad wings in profiles of all emission lines in the integrated spectrum (Fig. \ref{fig:J2048_s1fit_line}). 
The outflow wings dominate the line profiles of \oiiilong{} doublets with a flux fraction of 84\%.
The \oiii\ outflow have two components, 
a primary wing (``outflow line 1'' in Figure \ref{fig:J2048_s1fit_line}) with $v_\mathrm{s}$ of $-500\pm20$ \kms\ and FWHM of $1360\pm30$ \kms\ (Table \ref{tab:line_vel});
a secondary wing (``outflow line 2'') with $v_\mathrm{s}$ of $-2250\pm30$ \kms\ and FWHM of $1110\pm60$ \kms;
the line widths are corrected for spectral resolution. 
The primary \oiii\ outflow is marginally extended out of the PSF scale (Figure \ref{fig:J2048_image}, right panel), 
with a PSF-corrected effective radius of 1.4 kpc. 
The secondary wing traces a faster outflow component while the component is unresolved in this observation.  
The log(\oiii/\hb) ratio of the primary wing is 0.94$\pm$0.04 (Table \ref{tab:line_bpt}).
The \hb\ line of the secondary wing is not detected 
and we can derive a lower limit of log(\oiii/\hb), $\sim1.2$, 
by assuming the peak of the \hb\ line with the 1$\sigma$ noise level (Figure \ref{fig:J2048_s1fit_line}, bottom-left panel). 
Those high \oiii/\hb\ flux ratio and upper limit imply that the fast outflowing gas is ionized by AGN. 

We follow the method adopted in \cite{Chen2025} to estimate the mass-loss rate and kinetic power of the fast ionized outflow. 
The gas mass of the ionized outflow ($M_\mathrm{out}$) can be derived with the fluxes of outflow components of \oiiblong\ and \oiiiblong{} lines,
with the assumptions that the ionized outflow has a solar oxygen abundance
and the oxygen ions are mainly in the singly and doubly ionized phases. 
The extinction in the outflow, $A_{V,\mathrm{out}}=3.9\pm0.3$, is estimated using the Balmer decrement of the outflow lines. 
The electron density, $10^{3.1\pm0.8}$ \ccm{}, is estimated with the outflow component of the \sii\ doublets. 
The Oxygen-based\footnote{
	The $M_\mathrm{out}$ estimated from the flux of \ha\ outflow profile is $10^{8.1\pm0.8}$ \msun, 
	which is consistent to the Oxygen-based result. 
} $M_\mathrm{out}$ is then estimated to be $10^{8.0\pm0.8}$ \msun\ using the Equation (2) in \cite{Chen2025}. 
The timescale of the outflow can be derived as $\Delta t_\mathrm{out}=r_\mathrm{out} / v_\mathrm{out} = 0.7$ Myr;
here we adopt the effective radius, 1.4 kpc, as the outflow traveling distance.
The outflow velocity is calculated as $v_\mathrm{out}=|v_{50}|+w_{80}/2=2070\pm40$ \kms, 
where $v_{50}$ and $w_{80}$ are the 50\% velocity and the 80\% width of 
the sum of outflow 1 and 2 components of \oiii\ (Figure \ref{fig:J2048_s1fit_line}, bottom-left panel). 
The time-averaged mass-loss rate is estimated to be 
$\dot{M}_\mathrm{out}=M_\mathrm{out} v_\mathrm{out} / r_\mathrm{out} = 160$ \sfrunit.
The kinetic power (or the kinetic energy ejection rate) is estimated to be 
$\dot{E}_\mathrm{k,out}=\dot{M}_\mathrm{out} v_\mathrm{out}^2 /2 = 10^{44.3}$ \lumcgs.

The high velocity and kinetic power as well as the AGN-type ionization of the fast ionized outflow
suggest that it is driven by AGN activity. 
The coupling factor of kinetic power, $\dot{E}_\mathrm{k,out}/L_\mathrm{AGN,bol}=0.2\%$,  
is consistent with the values of luminous AGNs \citep[e.g.,][]{Fiore2017}. 
The high mass-loss and kinetic ejection rates
indicate that the powerful outflow could have an important role in the galaxy evolution, 
e.g., by redistributing the fueling gas in the central and outskirt regions. 


\begin{figure*}[!ht]
    \begin{center}
		\includegraphics[trim=0 0 -36 0, width=0.9\textwidth]{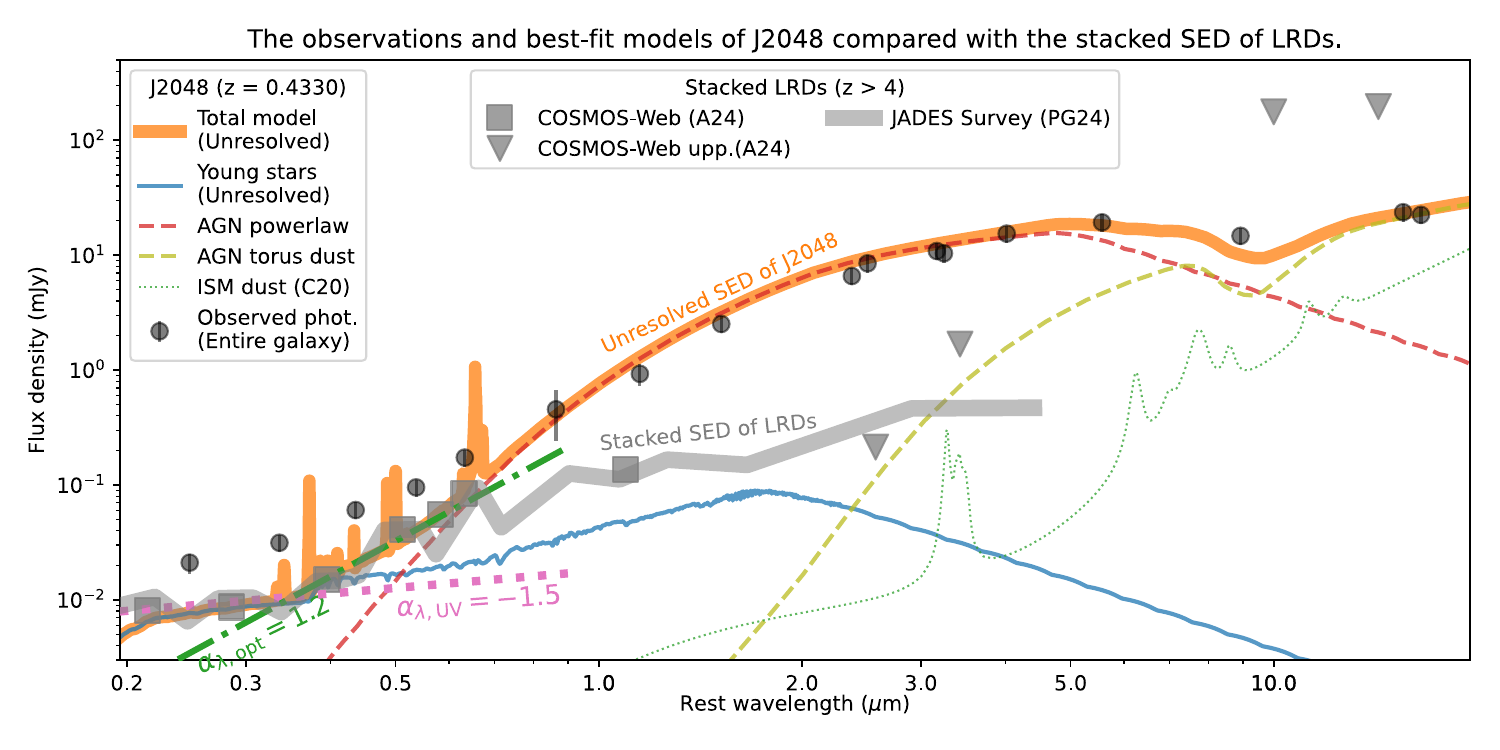}
    \end{center}
    \vspace{-20pt}
	\caption{
		Best-fit models of the simultaneous spectrum+SED fitting for J2048 in the unresolved ($r<2$\,kpc) region of J2048.
		The unresolved SED (orange thick line) is obtained by subtracting the best-fit models 
		of the extended outskirt region (Figure \ref{fig:J2048_outskirt_comp} and \ref{fig:J2048_outskirt_s1fit})
		from the best-fit SED of the entire galaxy (Figure \ref{fig:J2048_s3fit_SED}).
		The best-fit power-law models for the unresolved SED
		are shown in purple dotted (UV, $<3500$\,\AA) and green dash-dotted lines (optical, 4000--6500\,\AA), respectively.
		The stacked SED of LRDs in JADES \citep{perezgonzalez2024} and COSMOS-Web surveys \citep{akins2024}
		are shown in grey thick curves and squares, respectively;
		the upper limits of LRDs in COSMOS-Web survey are shown in grey triangles. 
		All stacked SED are normalized to 0.01 mJy at rest 3000\,\AA\ for a direct comparison of the unresolved SED of J2048 and high-$z$ LRDs. 
	}
	\label{fig:J2048_LRD_SED_comp}
\end{figure*}

\begin{figure*}[!ht]
    \begin{center}
	\includegraphics[trim=0 30pt 0 0, width=\textwidth]{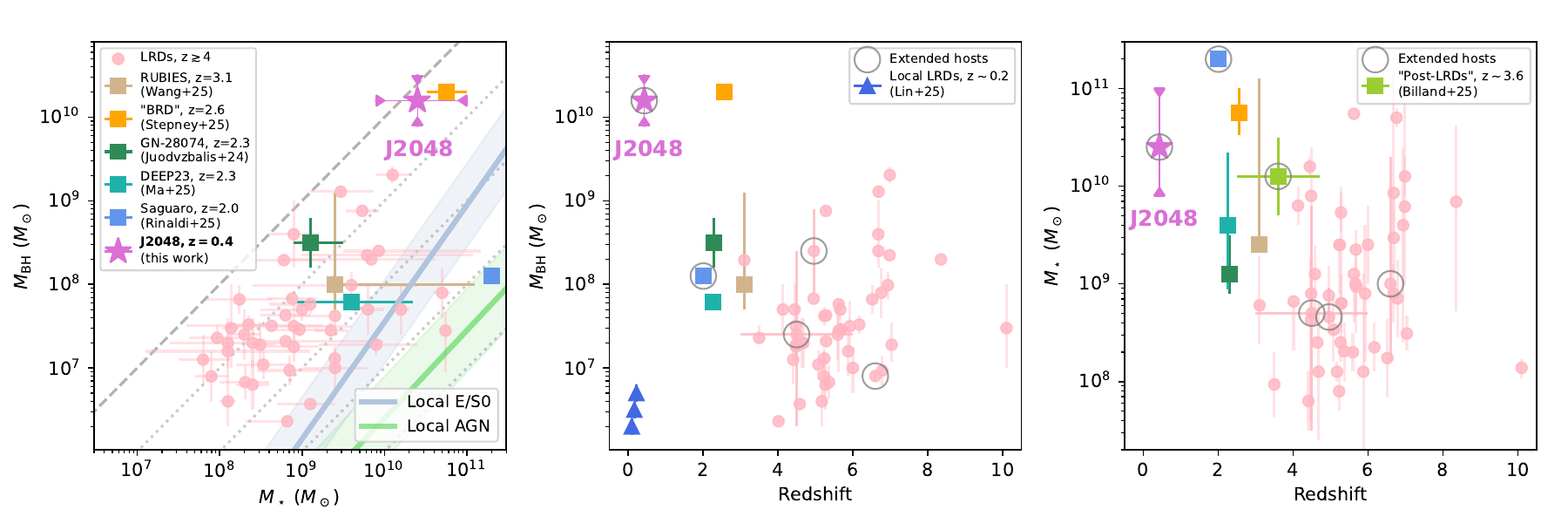}
    \end{center}
	\caption{
	$M_\mathrm{BH}$ vs. $M_\mathrm{*}$ (left), $M_\mathrm{BH}$ vs. redshift (middle), and $M_\mathrm{*}$ vs. redshift (right)
	diagrams of JWST-discovered LRDs at $z\gtrsim4$
	\citep[pink dots,][]{Goulding2023,Harikane2023,kocevski2023,maiolino2024,Wang2024,Juodzbalis2024,Akins2025,CChen2025LRD,Iani2025,kocevski2025,Zhuang2025}, 
	as well as the lower-$z$ LRD-like galaxies at Cosmic Noon \citep[squares,][]{Juodzbalis2024b,Billand2025,ma2025,Rinaldi2025,Stepney2024,wang2025}
	and local universe \citep[triangles,][]{Lin2025}. 
	J2048 is shown as purple stars with the lower- and upper-limits (Section \ref{subsec:result_agn} and \ref{subsec:result_host}) denoted in purple triangles. 
	In the left panel, the $M_\mathrm{BH}$-$M_\mathrm{*}$ relations of local AGNs and E/S0-type galaxies
	are shown with green and blue thick lines, respectively \citep[with $\pm$0.5 dex;][]{Reines2015}. 
	The $M_\mathrm{BH}/M_\mathrm{*}=100\%$ locations are shown with grey dashed line;
	the ratios of $0.01\%$--$10\%$ are shown with grey dotted lines. 
	In the middle and right panel, objects with extended host galaxies are denoted with open circles \citep{Billand2025,CChen2025LRD,Iani2025,Rinaldi2025,Zhuang2025}. 
	Note that all the $M_\mathrm{BH}$ shown in the plots are estimated assuming a Doppler broadening of hydrogen lines 
	except for \citet{Rinaldi2025} in which an X-ray calibration is adopted.
	} 
	\label{fig:MBH_Mstar_comp}
\end{figure*}

\section{Summary and discussion} \label{sec:discussion}

In this paper, we report a new discovery of a LRD-like galaxy at $z=0.4$, J2048, 
utilizing the new GMOS optical IFU spectroscopic data 
and the archived multi-band photometric SED.
J2048 has a v-shaped SED in UV and optical bands as those appear in JWST-discovered LRDs at $z>4$.
The blue-excess is spatially extended and emitted from a starburst with SFR of 400 \sfrunit.
A compact red continuum in NIR and the BLR \ha\ line indicate an obscured AGN with $A_{V,\mathrm{PL}}=6.3$, 
which corresponds to a bolometric luminosity of $10^{13.6}$ \lsun.
The black hole mass is estimated to be $10^{10.2}$ \msun. 
J2048 also exhibits an extended, star formation ionized narrow line region
and a concentrated, highly fast ionized outflow driven by the AGN, 
which suggest strong interaction between the AGN and the host galaxy. 

We discuss the observational similarities of J2048 and high-$z$ LRDs
in Section \ref{subsec:discuss_low_z_LRD}.
The implications on high-$z$ LRDs from the observations of J2048 
are discussed in Section \ref{subsec:discuss_implication}
and the overmassive SMBH in Section \ref{subsec:discuss_over_massive}. 

\subsection{A LRD-like galaxy at low-redshift} \label{subsec:discuss_low_z_LRD}

J2048 and the LRDs at $z>4$ share several common features:
(i) a v-shaped SED in the UV and optical bands (Figure \ref{fig:J2048_LRD_SED_comp});
(ii) a compact red continuum and a broad \ha\ line component
emitted by attenuated AGN (Figure \ref{fig:J2048_image} and \ref{fig:J2048_s3fit_SED}); 
and (iii) an overmassive location in the $M_\mathrm{BH}$-$M_\star$ diagram (Figure \ref{fig:MBH_Mstar_comp}). 
These similarities suggest J2048 as a low-$z$ analog of those distant LRDs. 

The SED of J2048 is compared with 
the stacked SED of high-$z$ LRDs in JADES \citep{perezgonzalez2024} and COSMOS-Web surveys \citep{akins2024}
in Figure \ref{fig:J2048_LRD_SED_comp}.
For the sake of a direct comparison with SED of high-$z$ LRDs that are typically unresolved,
we show the SED from the unresolved region ($r<2$\,kpc) of J2048 in Figure \ref{fig:J2048_LRD_SED_comp} (orange thick line), 
which is obtained by subtracting the best-fit models of the extended outskirt spectrum\footnote{
	In order to obtain the SED in the outskirt region, 
	we extend the best-fit stellar continuum model (as shown in Figure \ref{fig:J2048_outskirt_s1fit}) to UV and NIR ranges.
	Since the AGN power-law continuum and torus components emit only in the nuclear region, 
	they are not considered in the subtracting of outskirt SED. 
}
from the best-fit SED of the entire galaxy (Figure \ref{fig:J2048_s3fit_SED}).
The stacked LRD SED are normalized to 0.01 mJy at rest 3000\,\AA\ 
to compare with the unresolved SED of J2048. 
The unresolved SED of J2048 is curved
due to the Balmer break of a young stellar population formed in $\sim20$ Myr (from continuum fitting)
and the red color from the obscured AGN power-law.
The unresolved SED of J2048
is very similar to the v-shaped SED of the stacked LRD templates
in the UV and optical ranges. 
The approximate power-law indexes\footnote{
	Note that the $\alpha_\mathrm{\lambda,UV}$ and $\alpha_\mathrm{\lambda,opt}$ discussed here correspond to the model SED of the unresolved region, 
	which have slightly different values from those discussed in Section \ref{sec:intro} and Figure \ref{fig:J2048_outskirt_comp}
	that are based on the integrated spectrum of the entire galaxy. 
}
in UV ($<3500$\,\AA) and optical range (4000--6500\,\AA)
are estimated to be $\alpha_\mathrm{\lambda,UV}=-1.5$ and $\alpha_\mathrm{\lambda,opt}=1.2$, respectively, 
which are within the index ranges of high-$z$ LRDs, e.g., $-2.8<\alpha_\mathrm{\lambda,UV}<-0.4$ and $0<\alpha_\mathrm{\lambda,opt}<3$
\citep[Figure 3 of][]{kocevski2025}. 
The v-shaped SED with a red continuum is only detected in the unresolved region ($r<2$\,kpc) of J2048,
while the extended region shows a blue continuum emitted by young stars in the optical range (Figure \ref{fig:J2048_outskirt_comp} and \ref{fig:J2048_outskirt_s1fit}). 

The SED of J2048 has an obvious red bump in the NIR range ($>1$\,\micron). 
We examine the possible contributor of the red bump in Section \ref{subsec:method_exam}, 
and conclude that a reddened AGN power-law continuum is the only reasonable explanation. 
The reddened power-law continuum as well as the BLR \ha\ line 
indicate an obscured AGN embedded in J2048, 
which is also considered as a main population in the high-$z$ LRDs \citep[e.g.,][]{maiolino2024,greene2024,hviding2025}.
The SED of J2048 is redder than the stacked LRD SED (Figure \ref{fig:J2048_LRD_SED_comp}), 
which is probably due to the higher extinction of J2048, $A_{V,\mathrm{PL}}=6.3$, 
compared to the typical values of LRDs, e.g., $A_V=3$--4 \citep{kocevski2025,li2025}.

Utilizing the best-fit results of the pure spectral fitting and the simultaneous spectrum+SED fitting, 
we estimate $M_\mathrm{BH}$ and $M_\star$ as well as their lower- and upper-limits (Section \ref{subsec:result_agn} and \ref{subsec:result_host}),
which are shown in Figure \ref{fig:MBH_Mstar_comp}. 
The results of the JWST-discovered LRDs at $z\gtrsim4$
\citep[e.g.,][]{Goulding2023,Harikane2023,kocevski2023,maiolino2024,Wang2024,Juodzbalis2024,Akins2025,CChen2025LRD,Iani2025,kocevski2025,Zhuang2025}, 
as well as the lower-$z$ LRD-like galaxies at Cosmic Noon \citep[squares,][]{Juodzbalis2024b,Billand2025,ma2025,Rinaldi2025,Stepney2024,wang2025}
and in the local universe \citep[triangles,][]{Lin2025}
are also shown in Figure \ref{fig:MBH_Mstar_comp}.  
Despite the large uncertainties in estimation of host stellar mass\footnote{
	The typical method to estimate $M_\mathrm{*}$ with SED fitting
	can be affected by the choice of SED types, e.g., galaxy-only SED or AGN-host composite SED \citep[e.g.][]{Wang2024,wang2025,Billand2025}. 
	The derived $M_\mathrm{*}$ can also be underestimated due to the lack of the faint old stellar population, especially for the LRD-like galaxies at lower redshifts 
	(as discussed in Section \ref{subsec:result_host}). 
	For $M_\mathrm{*}$ estimated based on dynamical mass, the uncertainties can originate from the unconstrained size \citep[e.g.,][]{ma2025}
	and the complicated dynamical structure \citep[e.g.,][]{Banerji2021,Stepney2024}. 
}, 
there is an increasing trend of $M_\mathrm{*}$ from LRDs at $z>4$ ($10^8$--$10^{10}$\,\msun)
to the analogs at Cosmic Noon ($10^9$--$10^{11}$\,\msun), 
and J2048 locates at this increasing path towards lower redshift as shown in Figure \ref{fig:MBH_Mstar_comp} (right panel). 
On the other hand, similar to many of the distant LRDs, 
J2048 also possesses an overmassive SMBH
even if we adopt our most conservative estimation of $M_\mathrm{BH}$ or $M_\mathrm{*}$
(Figure \ref{fig:MBH_Mstar_comp}, left panel). 
We discuss 
the overmassive SMBH in details in Section \ref{subsec:discuss_over_massive}. 

Finally we note the main difference of J2048's spectral features compared to those of high-$z$ LRDs, 
i.e., no obvious absorption feature on AGN BLR lines in J2048. 
The absorption line can be associated to the dense gas environments
and super-Eddington accretion bursts in high-$z$ LRDs \citep[e.g.,][]{inayoshi2025}. 
However, the non-detection of such absorption lines in J2048's spectrum
does not imply the actual absence of those features,
since in the observation only \ha\ show BLR line
while the \ha\ absorption line could be merged 
by the complicated outflow profiles of \ha-\nii\ complex. 
Future NIR observations of Paschen lines are required to address if such dense gas absorption exist in J2048.  

\subsection{Implications on high-redshift LRDs from observation of J2048} \label{subsec:discuss_implication}

\begin{table*}[!ht]
	\caption{Summary of LRDs and LRD-like galaxies with extended host emissions.}
	\vspace{-4mm}
	\centering
	\begin{tabular}{c|ccccc}
		\hline
		\hline
		Objects & Method & Redshift & Radius\footnote{
		The wavelength ranges of the radius are all in rest frame.
		} (kpc) & Stellar mass (log \msun) & BH mass (log \msun) \\
		\hline
		``Virgil'' & Images & 6.631 & 0.49 (0.2--0.9\,\micron) & $\sim9.0$ (stellar-only model) & 6.90$\pm$0.05  \\
		\citep{Iani2025} & ~ & ~ & ~ & $\sim8.5$ (stellar+AGN) & ~  \\
		\hline
		MSAID38108 & Images & 4.96  & 0.66 (0.2--0.5\,\micron) & 8.66$\pm$0.24 & 8.40$\pm$0.50 \\	
		\citep{CChen2025LRD} & ~ & ~ & ~ & ~ & ~ \\	
		\hline
		\citet{Zhuang2025} & Images & 3.0--6.0 & 0.4--0.8 (0.2--0.4\,\micron) & 7.5-9.8 & 6.3-8.4 \\
		\hline
		\citet{Billand2025} & Images & 2.5--4.7 & 0.2--0.8 (0.3\,\micron) & 9.7--10.5 & - \\
		\hline
		``Saguaro''         & Images, slit-spec. & 2.015 & 5.8 (0.2\,\micron) & 11.3 & 8.12 \\
		\citep{Rinaldi2025} & ~                  &  ~    & 3.7 (1.5\,\micron) &  ~ &  ~ \\
		\hline
		J2048 (this work) & IFU-spec. & 0.433 & 3--7 (0.3--0.7\,\micron) & 10.4$\pm$0.6 & 10.2$\pm$0.3 \\
		\hline
	\end{tabular}
	\label{tab:extended_host}
\end{table*}

Thanks to its brightness and low redshift, 
J2048 serves as a unique laboratory to witness the properties of distant LRDs. 

One of the major mysteries on LRDs is the reason of the v-shaped SED, or in other word, the origin of the blue excess. 
Either a star-forming galaxy or scattered AGN light is considered as possible explanations of the blue excess \citep[e.g.][]{leung2024}. 
Several works reported extended features in tens of LRDs ($z>3$) with JWST's imaging observations at rest UV or optical blue wavelengths
\citep[e.g.,][]{Rinaldi2024,Billand2025,CChen2025LRD,Iani2025,Zhuang2025}, 
which support 
the host star-forming origin of the blue excess in the v-shaped SED. 
However, due to the distance of those high-$z$ LRDs, the extended features are usually very faint
and make spectroscopic identification difficult.
\citet{Rinaldi2025} reported a LRD-like nucleus in a disk galaxy, ``Saguaro'' at $z=2.015$, utilizing NIRSpec slit spectra
as well as HST and JWST imaging observations.  
At the same time, the current spectroscopic data of ``Saguaro'' is 
confined to the very central region of the source, 
e.g., only covering the nucleus and a very limited fraction of the host, 
with the distinction between the nucleus and the extended component inferred mainly from photometries.
The new GMOS IFU observation of J2048 covers the entire host galaxy, 
which clearly exhibit that the extended, blue continuum is emitted by young stars
with a SFR of 400 \sfrunit{} in the recent $\sim20$ Myr. 
This is the first time that a starburst 
host galaxy in a LRD-like object is spatially resolved with 
IFU spectroscopic identifications;
which also provides the lowest-$z$ case of a LRD-like galaxy with an extended host morphology.
The spatially resolved, spectroscopically identified extended hosts founded in ``Saguaro'' ($z\sim2$) and J2048 ($z\sim0.4$)
suggest that such systems, i.e., a LRD-like nucleus embedded in a star-forming galaxy, 
may not be isolated cases. 

The properties of the reported LRDs and LRD-like galaxies with extended host components are summarized in Table \ref{tab:extended_host}.
These objects are also highlighted with open circles in Figure \ref{fig:MBH_Mstar_comp}. 
The comparison between the objects with extended hosts and the parent LRD sample raises a question that, 
whether (1) these objects are a special population of LRDs, e.g., a more evolved phase with higher stellar mass;
or (2) extended (i.e., not point-like) feature is a general property of LRDs, it is hard to be detected at high-$z$ due to the depth of the imaging surveys.
As shown in Figure \ref{fig:MBH_Mstar_comp}, at $z>4$, LRDs with extended hosts possess a similar stellar mass to that of the parent LRD population;
while at $z<4$, the objects with extended hosts tend to show higher stellar mass, though the sample size is still limited.
This trend
matches the prediction of \cite{Billand2025}, i.e., 
the outskirt component could become more apparent as redshift decrease due to the formation of host galaxies.
On the other hand, 
\citet{Rinaldi2025} found that the outskirt of ``Saguaro'' can be disappeared with the current survey depths if it is moved from $z\sim2$ to $z\sim7$;
and the stacked images of 99 high-$z$ LRDs show significant extended emission in 2--3\, kpc scale
in rest UV wavelength ($\sim0.2$\,\micron).
A similar stacking analysis with 217 LRDs at $z\sim6.5$ also reveals extended emission from host galaxies with a typical scale of $\sim0.2$\,kpc 
in rest optical wavelength ($\sim0.6$\,\micron; \citealt{Zhang2025}).
The two stacking analyses imply that LRDs may commonly possess extended emission in UV and optical ranges,
i.e., similar to the individual objects detected with extended components (Table \ref{tab:extended_host});
while the extended components could be too faint in most cases to be detected by the current surveys with limited depths. 
These studies suggest that both of the above two scenarios (outskirt formation or survey depths)
could play a role in the detection of extended features of LRDs at different redshifts.
Distinguishment between the two scenarios is not the aim of this paper; 
while in either case, J2048 can provide an important low-$z$ template of LRD-like galaxies
to test the evolutionary scenarios of LRDs through different redshifts. 

Most LRDs at $z>4$ show weak or non-detections of rest MIR emissions, 
which implying the lack of dusty tori ($T\simeq 300$--1500 K) in the vicinity of the SMBHs 
\citep[e.g.,][]{leung2024,perezgonzalez2024,williams2024,akins2024,setton2025}.
A torus is also absent in the ``Big Red Dot'' at $z\sim2.5$ \citep{Stepney2024}.
A possible explanation is that the AGN is reddened by dust out of the nuclear torus scale
\citep[10--30 pc; e.g.,][]{nenkova2008,honig2019,nikutta2021,li2025,K.Chen2025}. 
The AGN torus component is detected in J2048.
However, if we redshift the SED of J2048 to $z>4$
and scale it to the flux level of LRDs at rest 3000\,\AA, 
the torus component of J2048 is also below the detection limits of the current LRD surveys
at rest MIR bands \citep[e.g.,][]{akins2024}, e.g., 
Herschel/PACS at 70 and 160\,\micron\ ($\sim10$\,\micron\ in rest frame), 
as shown in Figure \ref{fig:J2048_LRD_SED_comp}. 
As discussed in Section \ref{subsec:result_agn}, 
the torus-to-bolometric luminosity ratio of J2048, $\sim0.03$, is 
0.4 dex lower than that of typical normal AGNs \citep[e.g.,][]{Ichikawa2019_BASS}. 
The lower ratio suggest that
the torus in LRD-like objects could be thinner than that in typical AGNs,
i.e., a smaller covering factor, 
which reduces the absorbed and reemitted energy by torus and results in its faintness. 
The slim and faint torus scenario can be incorporated into the one with extended absorber
to explain the weak MIR features of LRDs. 

It is suggested that a powerful outflow can have an important role in the evolution of LRDs, 
e.g., expelling gas from the inner AGN region, enlarge the galaxy size, enhance or suppress star formation 
\citep[e.g.][]{Billand2025,wang2025}. 
Outflow features have been seen in a small number of LRDs with JWST/NIRSpec in both of emission lines
\citep[e.g., \oiii,][]{Cooper2025,DEugenio2025} and absorptions \citep[e.g., He I,][]{Juodzbalis2024b,wang2025}.
Such outflows have also been detected in lower-$z$ LRD-like galaxies
with unresolved observations \citep[e.g.][]{Stepney2024,Lin2025}. 
J2048 has the first spatially-resolved outflow detection
and the most powerful outflow among LRD-like objects. 
Since the extinction in the AGN central region is high ($A_V\sim6$) of J2048 and only \ha\ has the BLR line, 
the ionized outflow is clearly detected in the \oiii\ line profile without contamination of \hb\ BLR line. 
It is fast ($-2070$ \kms) and extended to 1~kpc scale,
showing high mass-loss rate ($160$ \sfrunit) and kinetic power ($10^{44.3}$ \lumcgs). 
The finding indicates that such powerful outflow could also occur in high-$z$ LRDs 
and significantly affect the evolution and environment of those distant galaxies. 

The extinction of AGN in J2048, $A_V=6.3$, 
is higher than the typical values of LRDs, e.g., $A_V=3$--4 \citep{kocevski2025,li2025}. 
The finding could suggest that LRDs with higher obscuration, or even a buried AGN (i.e., a type-2 SED), 
may also exist in high-$z$ universe;
and there could be a dynamical change of obscuration level 
driven by AGN feedback via strong radiation or powerful outflows. 
Those obscured/buried AGN population 
has not been found yet due to their faintness
and the selection upper limit of JWST observations at redshift up to 10. 
It is implicated that the actual AGN population related to the JWST-discovered distant LRDs
could possess even a higher number density than the values based on the current LRDs surveys. 

\subsection{An extremely overmassive SMBH} \label{subsec:discuss_over_massive}

J2048 has an extremely overmassive SMBH.
If we adopt the fiducial estimation, 
$M_\mathrm{BH}=10^{10.2}$ \msun\ and $M_\mathrm{*}=10^{10.4}$ \msun\ (Section \ref{subsec:result_agn} and \ref{subsec:result_host}), 
the $M_\mathrm{BH}/M_\mathrm{*}$ ratio is derived to be $\simeq60\%$, 
which is approximately two orders of magnitude higher than the ratio of local E/S0-type galaxies \citep{Reines2015}
with a similar $M_\mathrm{*}$ (Figure \ref{fig:MBH_Mstar_comp}, left panel).

Overmassive central BHs (e.g., $M_\mathrm{BH}/M_\mathrm{*}>10\%$) are found in a large fraction of the LRDs at $z>4$ \citep[e.g.,][]{inayoshi2024,maiolino2024,kocevski2025}. 
The origin of the overmassive BHs are still debated, possible explanations are:
overestimation of $M_\mathrm{BH}$ if the broad hydrogen lines are mainly broadened by non-Doppler broadening mechanisms\footnote{
	As discussed in Section \ref{subsec:result_agn_mbh}, 
	in the cases of non-Doppler broadening, the $M_\mathrm{BH}$ of J2048 can be estimated to be $10^{9.1}$ \msun\ 
	assuming an Eddington accretion ($\lambda_\mathrm{Edd}=1$; e.g., \citealt{Torralba2025}). 
	Even with this the most conservative estimation, 
	the SMBH of J2048 is still overmassive with $M_\mathrm{BH}/M_\mathrm{*} \simeq5\%$, 
	one order of magnitude higher than the ratio of local E/S0-type galaxies \citep{Reines2015}
	with a similar $M_\mathrm{*}$ (Figure \ref{fig:MBH_Mstar_comp}, left panel).
}
(e.g., scattering or turbulence; \citealt{Kokubo2024,Chang2025,Torralba2025,Rusakov2025}), 
underestimation of $M_\mathrm{*}$ in SED fitting due to different choices of SED types \citep[e.g.,][]{maiolino2024,Wang2024},
selection bias of LRD surveys favoring luminous and massive BHs \citep[e.g.,][]{LiJ2025},
the lag of star formation and less developed stellar bulge at these early epochs \citep[e.g.][]{maiolino2024,Juodzbalis2024},
or the super-Eddington accretion bursts \citep[e.g.][]{inayoshi2024,Juodzbalis2024}. 
Such overmassive SMBHs are also discovered in two LRD-like galaxies at cosmic noon, 
ULASJ2315+0143 \citep[][]{Stepney2024}
and JADES GN-28074 \citep{Juodzbalis2024b}. 
Overmassive SMBHs are very rare in local universe \citep[e.g., NGC 1277,][]{vandenBosch2012}.
J2048 would possess one of the highest  $M_\mathrm{BH}/M_\mathrm{*}$ ratio at low redshifts. 

In the case of J2048, the possibility of an underestimated $M_\mathrm{*}$
can be discarded because
J2048 has an overmassive SMBH
even if we adopt our most conservative $M_\mathrm{*}$ estimation. 
Galaxy mergers are usually considered to enhance the accretion of central SMBHs \cite[e.g.][]{Hopkins2008,Sturm2011,Veilleux2013,Toba2022,Yutani2022}.
However, the overmassive SMBH of J2048 is unlikely the direct production of the recent merger in past $\sim20$ Myr (from stellar continuum fitting). 
Even if we consider J2048 experienced 10 cycles of such merger events, 
the mean accretion rate required to construct $M_\mathrm{BH}$ of $\sim10^{10}$ \msun\ is still impossibly high (e.g., $\sim$10--100 \sfrunit). 
Therefore the overmassive SMBH could already be in place before the recent merger event(s). 
A possible scenario is that, J2048 was a successor of the most massive LRDs at $z>4$, e.g., CEERS 7902 \citep[$M_\mathrm{BH}=10^{9.3}$, \msun][]{kocevski2025}, 
which already built the bulk of the $M_\mathrm{BH}$ at early epochs (e.g., via super-Eddington accretion bursts), 
and then experienced several merger events during Cosmic Noon to possess the observed $M_\mathrm{BH}$. 
Meanwhile, unlike most LRDs that left from the overmassive-BH phase 
as the development of the host stellar components \citep[e.g.][]{maiolino2024,Billand2025}, 
the stellar build-up of J2048 was suppressed due to some reasons 
(e.g., powerful outflow feedback, or stripping-out of stellar envelope during past mergers), 
and as a result, the overmassive-BH is retained till the observed epoch. 

J2048 is currently in a growing period of both of the SMBH and the host galaxy
suggested by the AGN Eddington ratio of 0.14 and the SFR of 400 \sfrunit\ (optical/IR-based). 
However, this active growing phase is likely to be terminated soon 
since a powerful outflows have been launched (Section \ref{subsec:result_outflow}),
although currently there is no significant evidence of the occurrence of such strong negative feedback\footnote{
	The instantaneous SFR estimated from narrow \ha\ or \oii\ lines is 20\%--40\% of 
	the average SFR in the past 10--20 Myr (estimated from optical stellar continuum or far-IR dust emission; see Section \ref{subsec:result_host} and \ref{subsec:result_narrow}).
	However, the decreasing trend of SFR is not as significant as those reported in a galaxy with an in-action negative feedback, 
	where SFR decreased over one order of magnitude in past several Myr \citep{Chen2025}. 
}
\citep[e.g.][]{Chen2025}. 
Furthermore, the current $M_\mathrm{BH}$ is already comparable to those of the most massive SMBHs in local universe \citep[e.g.,][]{Kormendy2013,Reines2015}, 
and it is not probable that J2048 will experience any further significant SMBH growth. 
Moreover, the host build-up in J2048 is also constrained, which prevent it to migrate into the local $M_\mathrm{BH}$-$M_\mathrm{*}$ relation. 
A local elliptical galaxy with $M_\mathrm{BH}$ of $\sim10^{10}$ \msun\ has a $M_\mathrm{*}$ of $\sim10^{12}$ \msun. 
If J2048 could grow up to $M_\mathrm{*}=10^{12}$ \msun\ via star formation in the remaining cosmic time ($\sim4.6$ Gyr at $z=0.433$), 
an average SFR of $\sim200$ \sfrunit\ over Gyrs is required, which is impossibly high. 
On the other hand, there is also no potential mass budget for future merger assembling. 
The Illustris TNG100 simulation suggests a maximum separation of $\sim200$ kpc ($\sim35\arcsec$ at $z=0.433$) for galaxy merger occurring in 4 Gyr \citep{Chamberlain2024}.
However, there is no non-stellar objects within $35\arcsec$ from J2048 that reaches 1\% of the flux of J2048 (19.05 mag in $i$-band). 
To summarize, the evolutionary destination of J2048 could be a moderately enlarged stellar bulge
with a retained overmassive SMBH. 

\section*{Acknowledgments}
We appreciate the anonymous referee for the constructive suggestions on this paper. 
We kindly appreciate Zhengrong Li for providing the stacked SED of JWST-discovered LRDs. 
This work is supported by the Japan Society for the Promotion of Science (JSPS) KAKENHI Grant Number 25K17447 (X.C.), 
25K01043 (K.I.), 24K00670 (M.A.), 24K22894 (M.O.), 23H00131 (A.K.I.), and 25K07370 (T.K.).
K.I. acknowledges support from JST FOREST Program Grant Number JPMJFR2466 and from Inamori Research Grants. 
M.A. was supported by NAOJ ALMA Scientific Research Grant Code 2023-24A.
The observations were carried out within the framework of Subaru-Gemini time exchange program which is operated by the National Astronomical Observatory of Japan. We are honored and grateful for the opportunity of observing the Universe from Maunakea, which has the cultural, historical and natural significance in Hawaii.
Data analysis was carried out on the Multi-wavelength Data Analysis System operated by the Astronomy Data Center (ADC), National Astronomical Observatory of Japan.

\facilities{
	Gemini-N/GMOS, 
	SDSS, 2MASS, WISE, Spitzer
}

\software{
	Astropy \citep{Astropy},
	Gemini IRAF \citep{GeminiIRAF},
	Matplotlib \citep{Matplotlib},
	PyNeb \citep{Luridiana2015},
	\texttt{S$^3$Fit}\citep{S3Fit}
} 



\appendix
\restartappendixnumbering


\section{GMOS integrated spectra in the central and outskirt regions}
\label{sec:appendix_outskirt}

Figure \ref{fig:J2048_outskirt_s1fit} shows the GMOS integrated spectra in the central and outskirt regions.
The central spectrum is created using the average spectra of the central pixels 
and scaled with the PSF model.
The outskirt spectrum is integrated with the GMOS data cube after subtracting the PSF-scaled spectra in the central region. 
Here we employ the PSF model 
as the average of the normalized intensity maps of AGN power-law and BLR \ha\ line, 
which are obtained with the per-pixel spectral fitting. 
The PSF can be fit with a Gaussian profile with FWHM of $0.65\arcsec$ (3.70 kpc) as shown in Figure \ref{fig:J2048_image} (right panel).

Since the emission from AGN power-law continuum is highly obscured, 
the central spectrum at rest $\lambda<6000$\,\AA\ 
is dominated by stellar continuum emitted by young stars. 
In the outskirt region where the AGN core components are not detected, 
the full spectrum in dominated by stellar light. 
Please refer to Section \ref{subsec:result_host} for detailed discussions. 

\begin{figure*}[ht!]
    \begin{center}
		\includegraphics[trim=0 0 -36 0, width=0.7\textwidth]{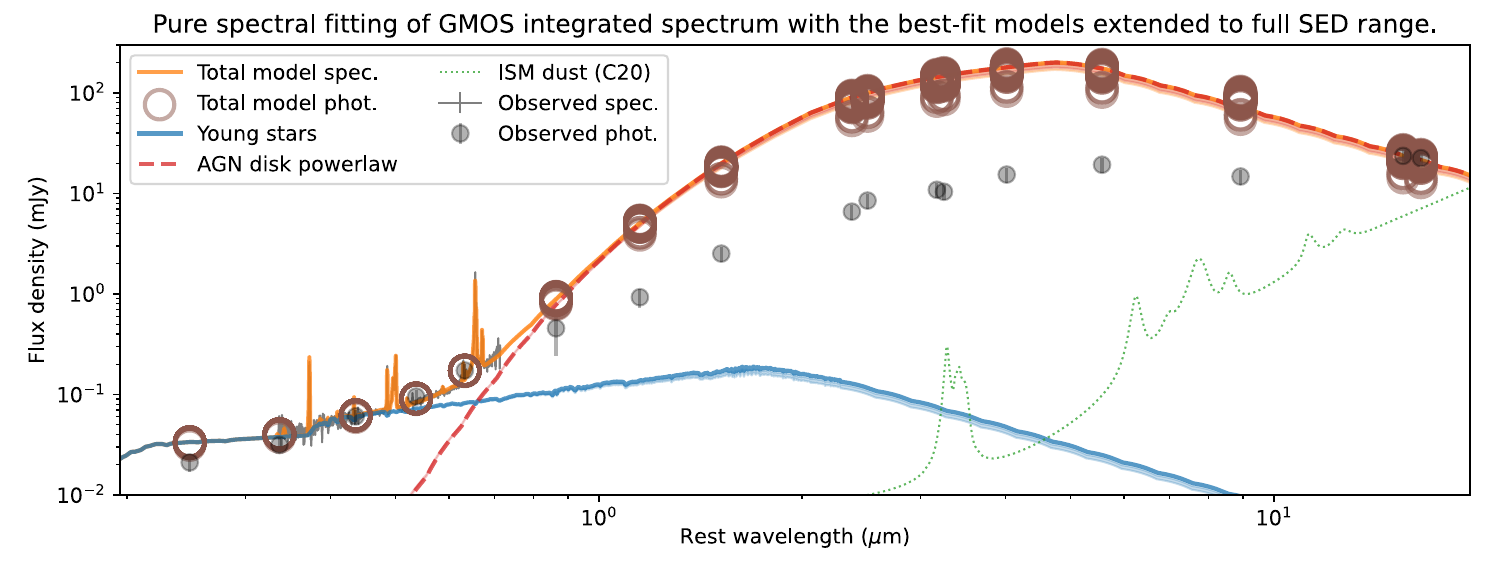}
    \end{center}
    \vspace{-20pt}
	\caption{
		Best-fit models of the pure spectral fitting for the GMOS integrated spectrum of the entire galaxy.
		The models are extended to the wavelength range covered by the photometric SED.
		The result suggests that the red power-law continuum is not well constrained with pure spectral fitting
		due to the limited wavelength coverage. See Section \ref{subsec:method_comp} for detailed discussion. 
	}
	\label{fig:J2048_s1fit_test_exSED}
\end{figure*}

\begin{figure*}[ht!]
    \begin{center}
		\includegraphics[trim=0 16 -36 0, clip, width=0.49\textwidth]{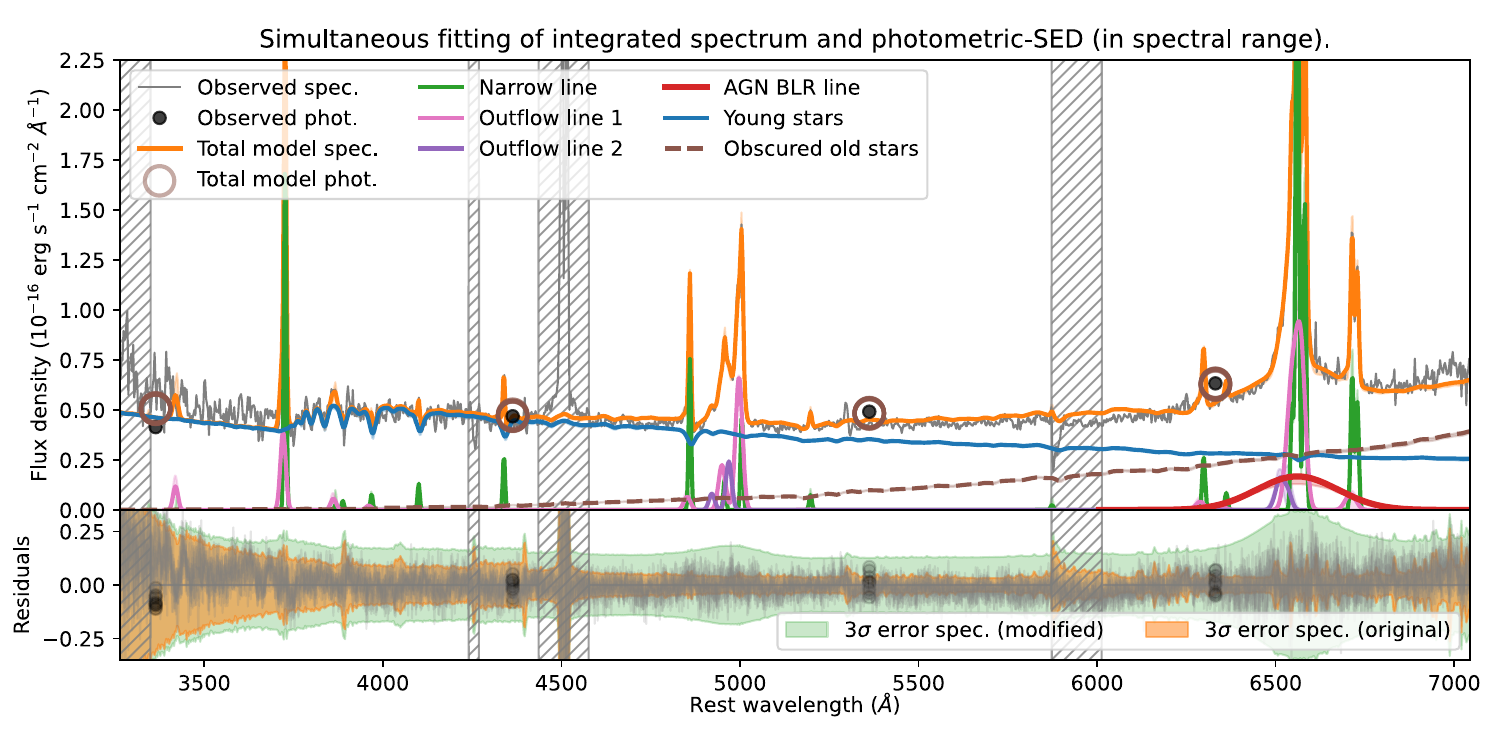}
		\includegraphics[trim=0 16 -36 0, clip, width=0.49\textwidth]{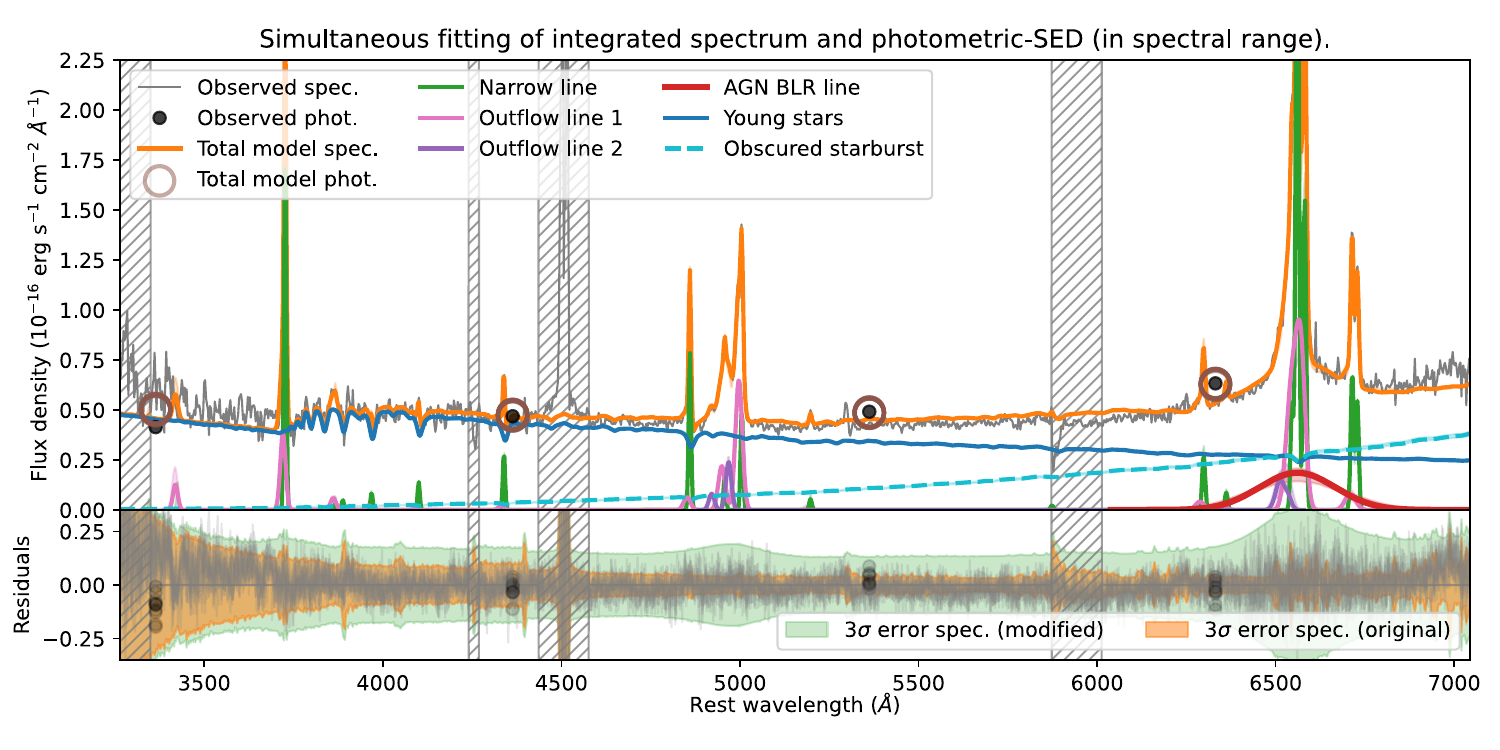}
		\includegraphics[trim=0 0 -36 12, clip, width=0.49\textwidth]{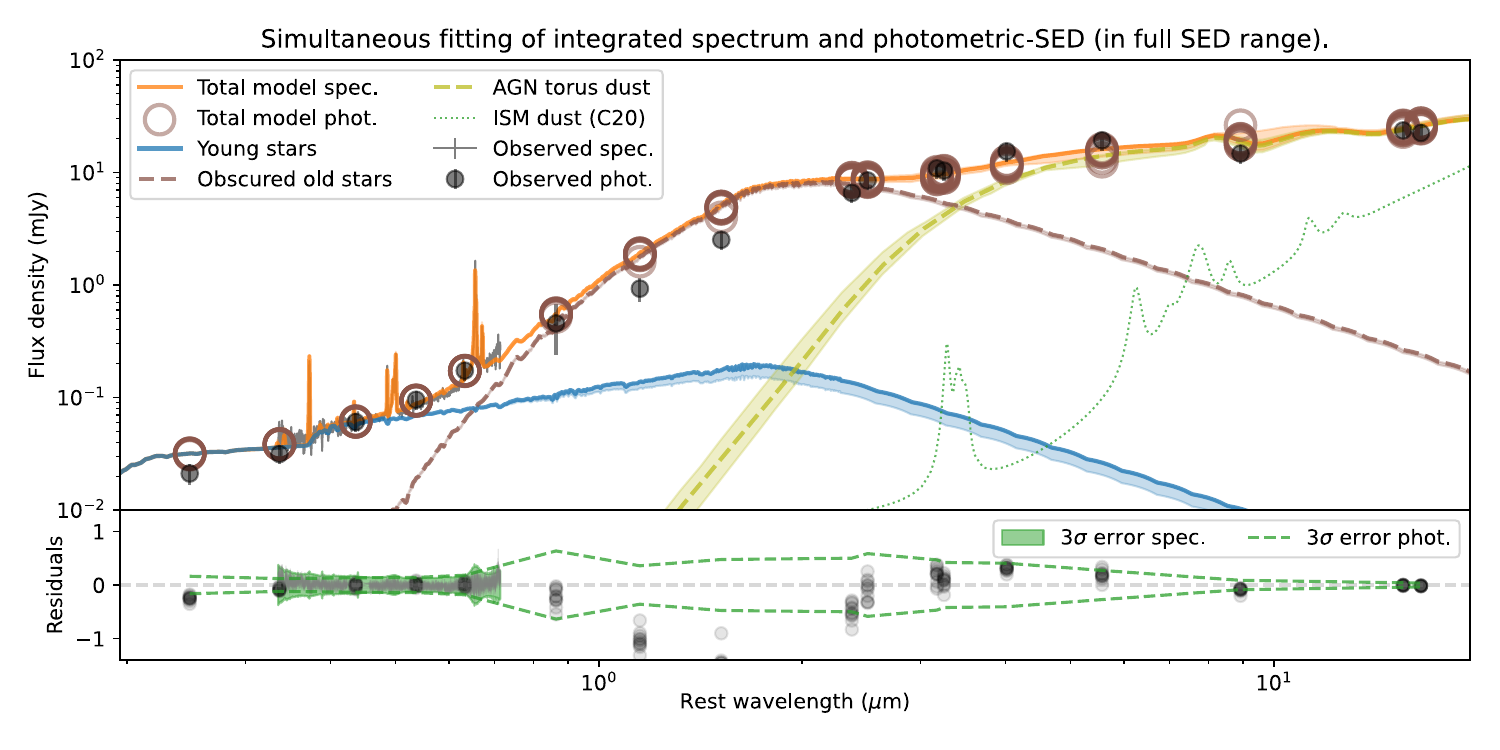}
		\includegraphics[trim=0 0 -36 12, clip, width=0.49\textwidth]{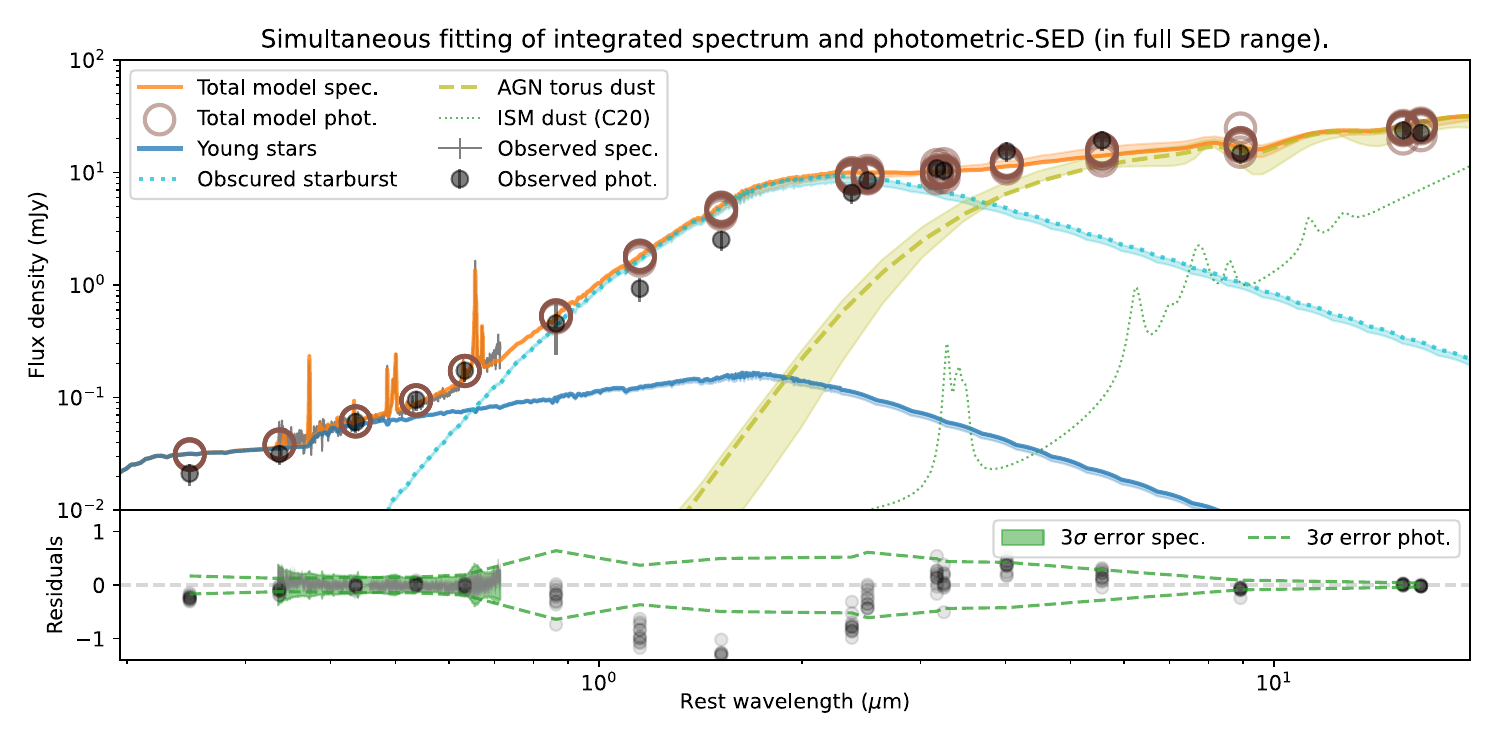}
    \end{center}
    \vspace{-20pt}
	\caption{
		Best-fit results of the simultaneous spectrum+SED fitting in the test cases using 
		(1) an obscured old stellar population (brown dashed curve in the left panels)
		or (2) a hidden starburst component (cyan dotted curves in the right panels) to reproduce the observed red bump in NIR SED. 
		See Section \ref{subsec:method_exam} for detailed discussion. 
	}
	\label{fig:J2048_s3fit_test_redSSP}
\end{figure*}

\begin{figure*}[ht!]
    \begin{center}
		\includegraphics[trim=0 20 -36   20, clip, width=0.7\textwidth]{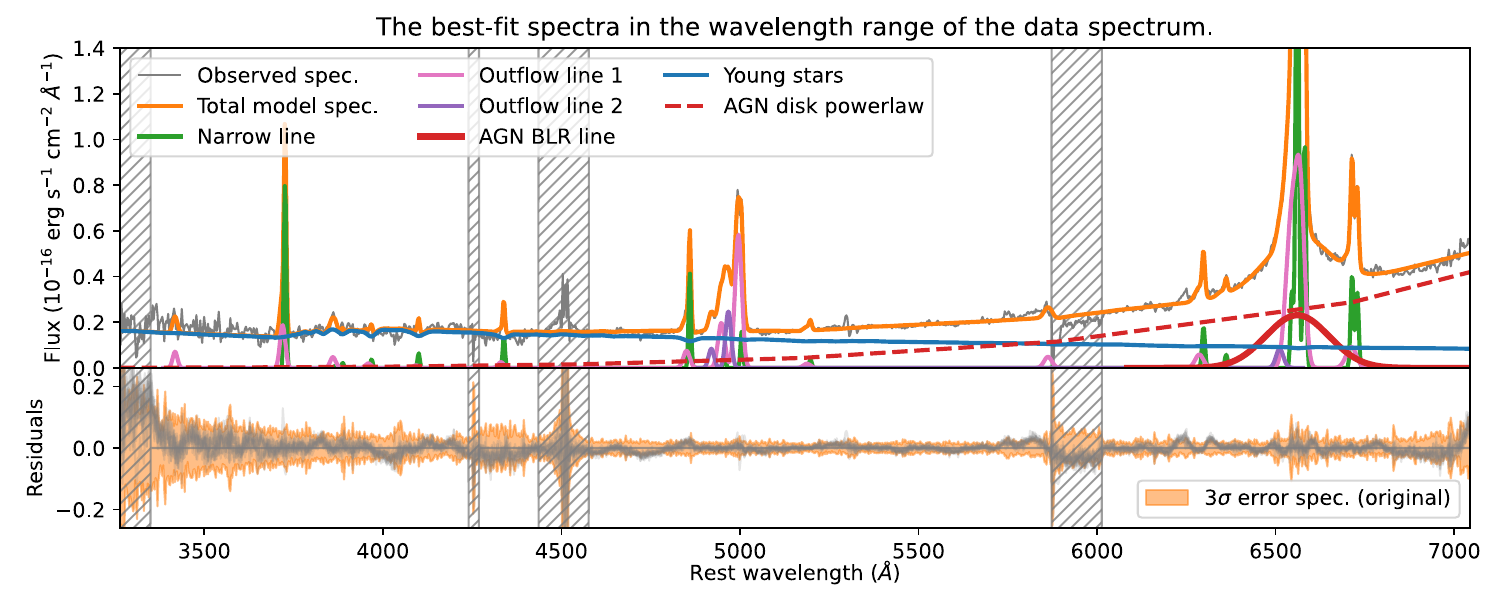}
		\includegraphics[trim=0  0 -36 19.5, clip, width=0.7\textwidth]{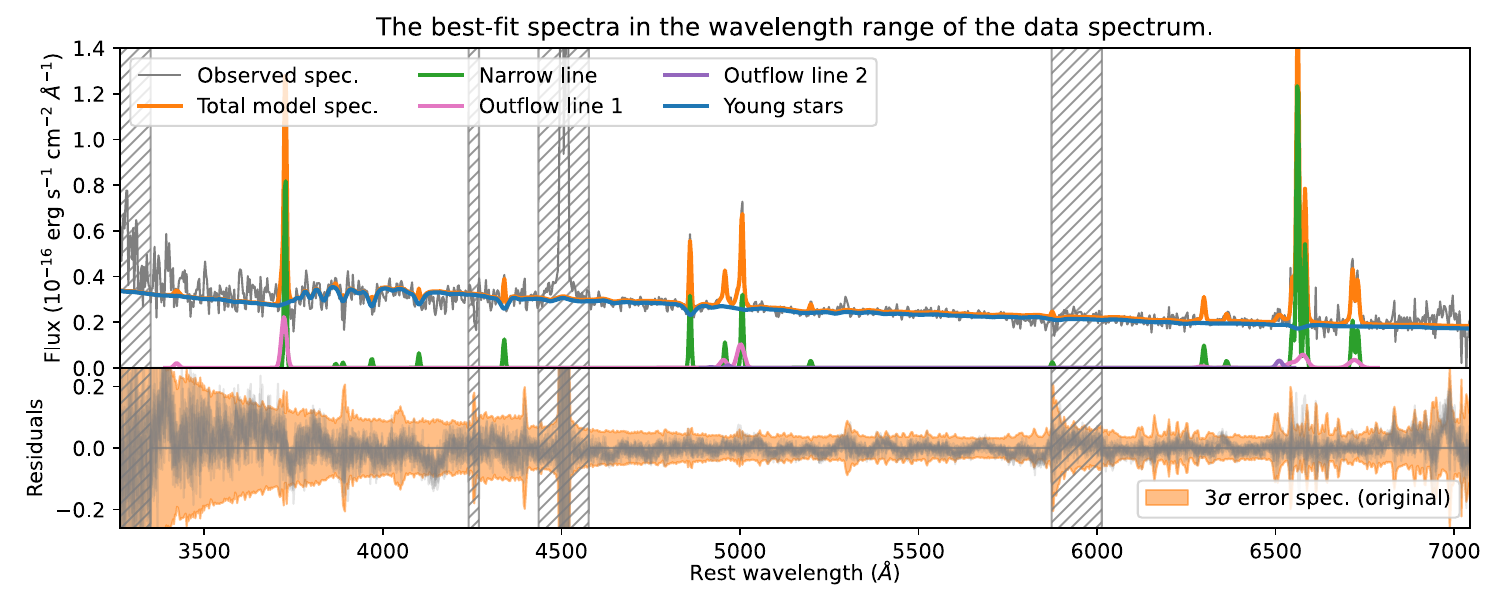}
    \end{center}
    \vspace{-20pt}
	\caption{
		GMOS spectra integrated in the central (PSF-scaled, top and upper-middle) and outskirt regions (lower-middle and bottom). 
		The best-fit models from the pure spectral fitting are shown with the same legends used in Figure \ref{fig:J2048_s1fit_line}. 
	}
	\label{fig:J2048_outskirt_s1fit}
\end{figure*}

\bibliography{low_z_LRD}{}
\bibliographystyle{aasjournal}



\end{document}